\documentclass[]{aa}  

\usepackage{graphicx}

\usepackage{subfig}
\usepackage{txfonts}
\usepackage{color}
\usepackage{lscape}
\usepackage{soul}
\usepackage{hyperref}
\hypersetup{draft}
\newcommand{\ugriz}{\textit{u', g', r', i', z'}}
\newcommand{\GALA}{\texttt{GALAPAGOS2}}
\newcommand{\galfitm}{\texttt{GALFIT-M}}
\newcommand{\z}{$z$}
\newcommand{\hst}{\textit{HST}}
\newcommand{\hstasc}{\textit{HST}-ACS}
\newcommand{\se}{\texttt{SExtractor}}
\newcommand{\photz}{$z_{\rm phot}$}
\newcommand{\morphsample}{{morphological sample}}

\newcommand{\otelodeep}{OTELO$_{deep}$}
\newcommand{\sersic}{$n$}
\newcommand{\sersicI}{$n_{I}$}
\newcommand{\sersicV}{$n_{V}$}

\newcommand{\ReGF}{$r_{\rm e}$}
\newcommand{\ReGFI}{$r_{{\rm e},I}$}

\newcommand{\N}{$N^{I}_{V}$}

\newcommand{\colourUR}{$({ u}\,-\,{ r})$}
\newcommand{\logmass}{$\log{M}_*/{M}_{\rm \odot}$}
\newcommand{\mstar}{$M_*$}
\newcommand{\msun}{$M_{\rm \odot}$}
\newcommand{\magiGF}{${I}_{\rm GF}$}
\newcommand{\magvGF}{${V}_{\rm GF}$}
\newcommand{\colourVIgf}{(${V}_{\rm GF}\,-\,{I}_{\rm GF}$)}
\newcommand{\concentration}{$c_{\rm 90/50}$}
\newcommand{\muoutput}{$\mu_{\rm output}$}

\begin{document}

   \title{The OTELO\ survey as a morphological probe.\\ Last ten Gyr of galaxy evolution.}
   \subtitle{The mass--size relation up to $z=2$.}
   \titlerunning{The OTELO survey as morphological probe.}
   
   \author{Jakub Nadolny\inst{1,2}
\and \'Angel Bongiovanni\inst{3,1,2,4}
\and Jordi Cepa\inst{1,2,4}
\and Miguel Cervi\~no\inst{5}
\and Ana Mar\'ia P\'erez Garc\'ia\inst{5,4}
\and Mirjana Povi\'c \inst{6,7}
\and Ricardo P\'erez Mart\'inez\inst{8,4}
\and Miguel S\'anchez-Portal\inst{3,4}
\and Jos\'e A. de Diego \inst{9}
\and Irene Pintos-Castro \inst{10}
\and Emilio Alfaro\inst{7}
\and H\'ector O. Casta\~neda \inst{11}$\dagger$
\and Jes\'us Gallego \inst{12}
\and J. Jes\'us Gonz\'alez \inst{9}
\and J. Ignacio Gonz\'alez-Serrano \inst{13,7}
\and Maritza A. Lara-L\'opez\inst{14} 
\and Carmen P. Padilla Torres \inst{1,15,16}
          }

   \institute{Instituto de Astrof\'isica de Canarias, E-38205 La Laguna, Tenerife, Spain 
          \and Universidad de La Laguna, Dept. Astrof\'isica, E-38206 La Laguna, Tenerife, Spain 
          \and Instituto de Radioastronom\'ia Milim\'etrica (IRAM), Av. Divina Pastora 7, N\'ucleo Central, E-18012 Granada, Spain 
          \and Asociaci\'on Astrof\'isica para la Promoci\'on de la Investigaci\'on, Instrumentaci\'on y su Desarrollo, ASPID, E-38205 La Laguna, Tenerife, Spain 
          \and Depto. Astrofísica, Centro de Astrobiología (INTA-CSIC), ESAC Campus, Camino Bajo del Castillo s/n, 28692, Villanueva de la Cañada, Spain 
          \and Ethiopian Space Science and Technology Institute (ESSTI), Entoto Observatory and Research Center (EORC), Astronomy and Astrophysics Research Division, PO Box 33679, Addis Abbaba, Ethiopia 
          \and Instituto de Astrof\'isica de Andaluc\'ia, CSIC, E-18080, Granada, Spain  
          \and ISDEFE for European Space Astronomy Centre (ESAC)/ESA, P.O. Box 78, E-28690, Villanueva de la Ca\~nada, Madrid, Spain 
          \and Instituto de Astronom\'ia, Universidad Nacional Aut\'onoma de M\'exico, Apdo. Postal 70-264, 04510 Ciudad de M\'exico, Mexico 
          \and Department of Astronomy \& Astrophysics, University of Toronto, Canada 
          \and Departamento de F\'isica, Escuela Superior de F\'isica y Matem\'aticas, Instituto Polit\'ecnico Nacional, M\'exico D.F., Mexico 
          \and Departamento de F\'isica de la Tierra y Astrof\'isica, Facultad CC F\'isicas, Instituto de F\'isica de Part\'iculas y del Cosmos, IPARCOS, Universidad Complutense de Madrid 
          \and Instituto de F\'isica de Cantabria (CSIC-Universidad de Cantabria), E-39005 Santander, Spain 
          \and Armagh Observatory and Planetarium, College Hill, Armagh BT61 9DG, Northern Ireland, UK 
          \and Fundaci\'on Galileo Galilei - INAF
          Rambla Jos\'e Ana Fern\'endez P\'erez, 7, 38712 Bre\~na Baja, TF - Spain 
          \and Telescopio Nazionale Galileo Roque de Los Muchachos Astronomical Observatory 38787 Garafia, TF, Spain 
          \\
                        \email{jnadolny@iac.es; quba.nadolny@gmail.com}
             }

   \date{Received ---; accepted ---}


  \abstract
  {The morphology of galaxies provide us with a unique tool for relating and understanding other physical properties and their changes over the course of cosmic time. It is only recently that we have been afforded access to a wealth of data for an unprecedented number galaxies thanks to large and deep surveys,
} 
  {We present the morphological catalogue of the OTELO survey galaxies detected with the \textit{Hubble Space Telescope} (\hst)-ACS F814W images. We explore various methods applied in previous works to separate early-type (ET) and late-type (LT) galaxies classified via spectral energy distribution (SED) fittings using galaxy templates. Together with this article, we are releasing a catalogue containing the main morphological parameters in the F606W and F814W bands derived for more than 8\,000 sources.}
   {The morphological analysis is based on the single-S\'ersic profile fit. We used the \GALA\ software to provide multi-wavelength morphological parameters fitted simultaneously in two \hstasc\ bands. The \GALA\ software detects, prepares guess values {for \galfitm}, and provides the best-fitting single-S\'ersic model in both bands for each source. Stellar masses were estimated using synthetic rest-frame magnitudes recovered from SED fittings of galaxy templates. The morphological catalogue is complemented with concentration indexes from a separate \se\ dual, high dynamical range mode.}
   {A total of 8\,812 sources were successfully fitted with single-S\'ersic profiles. The analysis of  a carefully selected sample of $\sim$\,3\,000 sources up to \photz\,=\,2 is presented in this work, of which 873 sources were not detected in previous studies. We found no statistical evidence for the evolution of the {low-mass end of} mass-size relation for ET and LT since \z\,=\,2. {Furthermore, we found a good agreement for the median size evolution for ET and LT galaxies, for a given stellar mass, with the data from the literature.} Compared to previous works on faint field galaxies, we found similarities regarding their rest-frame colours as well as the S\'ersic and concentration indices.}
   {}


   \maketitle
%

\section{Introduction}
\label{sec:introduction}

\begin{figure}[t!]
        \centering
        \includegraphics[width=\linewidth]{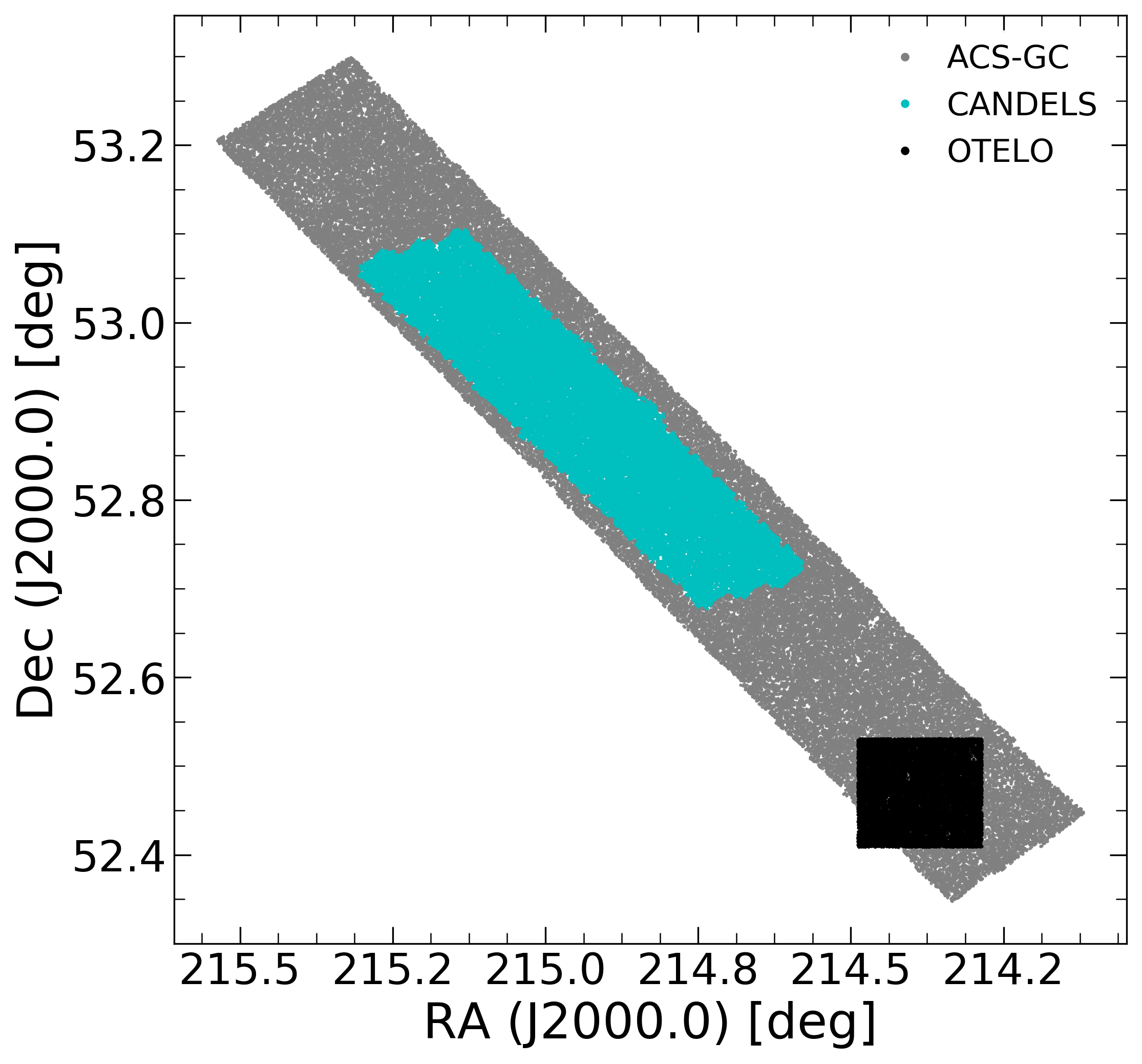}
        \caption[]{Spatial distribution of the morphological data available in EGS field. Data from ACS-GC \citep{Griffith2012ApJS..200....9G}, CANDELS \citep{CANDELS_AEGIS2012ApJS..200...13B}, and the OTELO survey \citep{OteloI}, are shown in gray, cyan, and black, respectively.
        \label{fig:egs_field}}
\end{figure}

Galaxy morphology is related to other physical properties such as star formation, dynamical histories, stellar mass, colours, luminosities, and different morphological parameters \citep{Kauffmann2003MNRAS.341...54K, Baldry2004ApJ...600..681B, Salim2007ApJS..173..267S,Povic2013MNRAS.435.3444P,Povic2012A&A...541A.118P, Schawinski2014MNRAS.440..889S, Mahoro2019MNRAS.485..452M}. Since \cite{Hubble1925ApJ....62..409H} discovered the extragalactic nature of some 'nebulae', the first step for tackling their systematic study was to establish a morphological classification. \cite{Hubble1936rene.book.....H} proposed the widely known \textit{tuning-fork} classification scheme, which was further extended by de Vaucouleurs system \citep{Vaucouleurs1959HDP....53..275D}, which incorporated the numerical stages as well. Soon it was realized that ellipticals and spirals had different photometric and dynamical properties. Ellipticals  were usually defined by red colours, a de Vaucouleurs radial profile (\citealt{Kormendy1977ApJ...218..333K}), and mainly sustained by the velocity dispersion; whereas spirals were bluer, with a radial profile composed of a de Vaucouleurs plus an exponential disk \citep{Freeman1970ApJ...160..811F} and they were dynamically sustained by rotation velocities. As a consequence, elliptical galaxies were compliant with a 'fundamental plane' \citep{Dressler1987ApJ...313L..37D}, whereas spirals obey a Tully-Fisher relation \citep{TF_RELATION_1977A&A....54..661T}. Furthermore, morphology is directly linked to the physical and chemical properties of galaxies, and, ultimately, to their evolution. In particular, {the stellar mass--size} relation (MSR) has been studied for several decades, and it has been shown that the galaxy size not only can vary significantly with stellar mass and morphology, but it also evolves with redshift depending on these two parameters \citep[e.g.][]{Kormendy1996ApJ...464L.119K,Shen2003MNRAS.343..978S,vanderWel2014ApJ...788...28V,LangeGAMA_Mass_Re_2015,Roy2018,Mowla2019ApJ...880...57M}. For instance, early-type (ET) galaxies are typically more compact for the same mass, and more massive for the same size, than the late-type (LT) galaxies at given redshift. These differences reflect different processes of the size growth in time \citep[see][and references therein]{Mowla2019ApJ...880...57M}. 

The morphological classifications of large galaxy collections at different redshifts provided by the rich surveys of recent decades (e.g. SDSS, \citealt{York2000AJ....120.1579Y}; VVDS, \citealt{VVDS2005A&A...439..845L}; COSMOS, \citealt{Cosmos2007ApJS..172....1S}) have required the use of automatic classification tools \citep[e.g.][]{Barden2012MNRAS.422..449B,Strateva2001AJ....122.1861S} and machine learning techniques (e.g. \citealt{Huertas2015ApJS..221....8H}).
After analysing a sample of $\sim$150\,000 galaxies from SDSS, \citet{Strateva2001AJ....122.1861S} have shown that using the \colourUR\ colour, it is possible to separate ET from LT galaxies. Later, more sophisticated methods were developed. Among these approaches, two  stand out in particular: non-parametric \citep{Abraham1994ApJ...432...75A,Abraham1996ApJS..107....1A, Abraham2003ApJ...588..218A, Bershady2000AJ....119.2645B, Conselice2000ApJ...529..886C, Lotz2004AJ....128..163L, Huertas2008A&A...478..971H, Povic2013MNRAS.435.3444P,Povic2015MNRAS.453.1644P} and parametric based on physical \citep{Galfit2002,GALFIT_2010AJ....139.2097P, Simard2002ApJS..142....1S, Simard2011ApJS..196...11S, deSouza2004ApJS..153..411D, Barden2012MNRAS.422..449B} or mathematical galaxy parameters \citep{Kelly_2005, Ngan2009MNRAS.396.1211N, Andrae2011MNRAS.417.2465A,Andrae2011MNRAS.411..385A,JimenezTeja2012ApJ...745..150J}. Both methods greatly depend  on the sensitivity of the data used \citep[e.g.][and references therein]{Haussler2007,Povic2015MNRAS.453.1644P}. 

The main advantage of the non-parametric methods is that they do not depend on any analytic form a priori and the information that is used is obtained directly from the source images (i.e. concentration index, colour, asymmetry, Gini index, smoothness, etc.). On the other hand, the main benefit of using a particular parametric function is that it can be extrapolated in the low signal-to-noise (S/N) source and can account for light at large radii
\citep{Haussler2007,galapagos2013MNRAS.430..330H}. A commonly used parametric form is a \cite{Sersic1968adga.book.....S} profile defined as:
\begin{equation}
\label{eq:sersic}
I (r) = I_{e} \exp \{-b_{n}[(r/r_{e})^{1/n} - 1]\},
\end{equation}
where $r_{e}$ is the effective radius (i.e. radius containing 50\% of total flux), $I_{e}$ is the intensity at $r_{e}$, \sersic\ is the S\'ersic index, and $b_{n}$ is a function of \sersic\ as defined in \cite{Ciotti1991A&A...249...99C}. Even through it may be less flexible than non-parametric methods, and assuming that the chosen model correctly describes the light distribution, this method is considered to be sufficiently robust  and feasible. In the particular case of the Sérsic profile, it has been used with success in previous studies \citep{Simard1998ASPC..145..108S,Graham2005AJ....130.1535G,Haussler2007,galapagos2013MNRAS.430..330H}. 

The aim of this study is to complement the OTELO survey \citep{OteloI} data-base\footnote{\url{http://resaerch.iac.es/projecto/otelo}} with a morphological analysis of the counterparts detected in high-resolution \hstasc\ F814W image up to $z\,=\,2$. The OTELO (OSIRIS Tunable Filter Emission Line Object) is a very deep, blind spectroscopic survey centered in a selected region of Extended Groth Strip (EGS). The morphological analysis is important for a full exploration of the possible scientific cases of the OTELO survey. Among them, we can highlight the following: a census of ET galaxies with emission lines, a comprehensive study of the properties of compact galaxies, a comparison of extragalactic sources with and without detection of emission lines; and a recently published work on the machine learning
techniques to separate ET from LT galaxies \citep{JAD}. { Additionally, in this work, we study the MSR of galaxies up to $z=2$ and down to stellar masses of \logmass$\,\sim\, 8$, which is $\sim\,$1 dex lower than the lower mass limit established in previous studies at the same redshift range \citep{vanderWel2014ApJ...788...28V,Mowla2019ApJ...880...57M}. In particular, we present insights on the median size evolution $r_{\rm e}$--$z$\, for LT and ET galaxies at a fixed stellar mass found in the OTELO field. } 

While there are few morphological studies in the EGS field \citep[][]{Povic2009ApJ...706..810P,Griffith2012ApJS..200....9G,CANDELS_AEGIS2012ApJS..200...13B}, the only overlapping with the OTELO survey field can be found in the work of \citet[][the ACS-GC catalogue, see Figure \ref{fig:egs_field}]{Griffith2012ApJS..200....9G}. The depth of the OTELO survey (27.8 AB on the \otelodeep\ image, see \citealt{OteloI}) and the stellar mass range (down to \logmass$\,\sim\,6$, see \citealt{Nadolny}) of the galaxies observed are the principal motivations for re-processing the archival \hstasc\ data, rather than employing morphological information from the previously published ACS-GC catalogue from \citet{Griffith2012ApJS..200....9G}. Furthermore, setting it in comparison with the ACS-GC catalogue, we analyse the \hstasc\ data using newer version of the {\tt GALFIT} software \citep{Galfit2002,GALFIT_2010AJ....139.2097P}; namely, its multi-wavelength version, the \galfitm\ \citep{galapagos2013MNRAS.430..330H}, where two \hstasc\ bands are analysed simultaneously. This multi-wavelength version has been shown to produce more accurate, complete, and meaningful results, especially in the low signal-to-noise (S/N) regime (\citealt{galapagos2013MNRAS.430..330H}; see also Sec. \ref{sec:comparision_with_griffith}). Extensive simulations in \citet{Haussler2007} have shown that \texttt{GALFIT} is quite robust and effective. Furthermore, it has been successfully used in several low- and high-redshift studies of surveys  \citep[e.g.][]{Barden2005ApJ...635..959B,Krywult2017A&A...598A.120K}, in different environment \citep[e.g.][]{Kuchner2017A&A...604A..54K}, as well as its multi-wavelength version \galfitm\ \citep[e.g.][]{Vulcani_MegaMorph_2014,Vika_MegaMorph2015}.

This paper is organised as follows. In Section \ref{sec:data}, we describe the data used in this work. In Section \ref{sec:methods}, we present the description of the method used in this work. Section \ref{sec:sample_selection} provides the sample selection process and quantitative comparison with \citet[][]{Griffith2012ApJS..200....9G}. In Section \ref{sec:morphological_analysis}, we analyse the results of our fitting process and compare it to the ET and LT classification based on SED fitting. {Furthermore, we analyse the MSR and compare it with previous works (Sect. \ref{sec:morphological_analysis_msr})}. In Section \ref{sec:discussion}, we provide a discussion of (i) the results of our morphological analysis using the \colourUR\ and compared it to the work of \citet{Strateva2001AJ....122.1861S}, as well as (ii) a discussion of the MSR found in this work {and median size evolution for a given stellar mass since $z=2$}. In Sections \ref{sec:morphological_catalogue} and \ref{sec:conclusions}, we present a description of the morphological catalogue and our conclusions, respectively. When necessary we adopt the cosmology with $\Omega_{\Lambda}=0.69$, $\Omega_{m}=0.31$, and H$_{0}=67.8$ km s$^{-1}$ Mpc$^{-1}$ from \cite{Planck2016A&A...594A..13P}.

\section{Data}
\label{sec:data}
\subsection{The OTELO survey}
\label{sec:otelo_catalogue}

The OTELO survey is based on the red tunable filters (RTF) images of the OSIRIS \citep{2003SPIE.4841.1739C} instrument at Gran Telescopio Canarias. A total of 36 RTF tomography slices were obtained. 
These slices were uniformly distributed in the spectral range between 9070 \AA\ and 9280 \AA, centered at 9175 \AA. Following their reduction and alignment, the co-addition of these slices provide a detection image \otelodeep,\ from which a total of 11\,273 raw detections were extracted. 
{ The \otelodeep\ photometry was complemented with archival data that was reprocessed, and PSF-matched to the OTELO's resolution (pixel scale of 0.254 arcsec/px),  available in EGS. Altogether, these data allowed to build the OTELO photometric catalogue\footnote{For the extensive list of archival data included in the OTELO photometric catalogue we refer to \citet[][]{OteloI}} }. 
Among the reprocessed archival data to OTELO pixel resolution are the images from \hstasc\ (F606W and F814W) and from the Canada-France-Hawaii Telescope Legacy Survey (CFHTLS\footnote{\url{https://www.cfht.hawaii.edu/Science/CFHTLS/}}; in \ugriz\ filters). Furthermore, the CFHTLS D3-25 $i'$-band source catalogue was used as the OTELO's astrometric reference. The OTELO photometric catalogue was used to estimate photometric redshift \photz\ with \texttt{LePhare} code \citep{Arnouts1999, Ilbert2006}. Each photometric redshift solution obtained with \texttt{LePhare} code is associated with a specific galaxy template, which corresponds to a particular galaxy type. Templates used in this work are the following: four Hubble-type templates (E, Sbc, Scd, Im) from \cite{Coleman1980} and six starburst galaxy templates from \cite{Kinney1996}. For the purpose of this work we consider those sources with a best-fitted template of the Hubble type E as ET , while LT refers to the sources fitted with the remaining ones.

The stellar masses were estimated following the method from \cite{LopezSanjuan2018}, using rest-frame synthetic magnitudes obtained from the templates described above. For further details on the stellar masses \mstar\ and physical size estimations used in this work, we refer to \citet{Nadolny}.
We refer to the data-products described in this section as low-resolution (or low-res) because these were obtained on the basis of the RTF images with pixel scale of 0.254"/px. On the other hand, the morphological parameters were obtained from the original high-resolution \hstasc\ images with pixel scale of 0.03 "/px.

\subsection{High-resolution \textit{Hubble Space Telescope} data}
\label{sec:astrometry}
The morphological catalogue is based on high-resolution images from \hstasc\ F606W and F814W filters (hereafter $V$- and $I$-band, respectively). A total of 11 (overlapping with the OTELO survey field) \hstasc\ tiles were retrieved from All-sky Extended Groth Strip International Survey (AEGIS) database\footnote{ \url{http://aegis.ucolick.org/mosaic\_page.htm}} in its native pixel scale of 0.03 arcsec/px and average resolution of $\sim$\,0.15 arcsec. This high-resolution data were acquired in HST Cycle 13 GO program 10134 (PI: M. Davis). In short, the \hstasc\ images were already processed with standard ACS pipe-line (including bias subtraction, gain, and flat-field correction) and a python-based multi-drizzle package \citep{Koekemoer2003} was applied to combine all exposures in one tile (including registration, median image creation, the identification and removal of cosmic rays).

 In this study, we aim to complement the OTELO survey with the morphological analysis, thus we have to align the \hstasc\ images to the same reference catalogue as used in the OTELO survey, that is, to the CFHTLS D3-25 $i'$-band source catalogue. This catalogue have an internal root mean square (RMS) astrometric error of 0.064 and 0.063 arcsec in equatorial coordinates\footnote{See "T0007 : The Final CFHTLS Release" at  \url{http://terapix.iap.fr/cplt/T0007/doc/T0007-doc.pdf}}. An astrometry correction is necessary not only to provide homogeneous celestial coordinates of objects detected on the \otelodeep\ image and its counterparts from high-resolution data, but also in our further analysis of morphological parameters. The initial astrometric offset between all of the 11 tiles used and CFHTLS catalogue were found to be different. Thus, we decided to provide a homogeneous astrometric correction for each of the tiles separately to the CFHTLS D3-25 $i'$-band catalogue before proceeding to the mosaic assembly.

The selection of the objects suitable for astrometry correction is important because we are matching low-resolution CFHTLS catalog (resolution $\sim$\,0.6 arcsec with pixel scale of 0.186 arcsec/px) with high-resolution \hstasc\ data (with resolution of $\sim$\,0.15 arcsec and pixel scale of 0.03 arcsec/px). This selection was based on (i) CFHTLS $i'$-band magnitude ($\leq$\,24.5), (ii) \se\  (\citealt{SEx1996}) parameters \texttt{CLASS\_STAR}\,$\geq$\,0.9 (for both images: \hst\ $I$-band and CFHTLS $i'$-band), and (iii) \texttt{FLAG}\,=\,0 for only high-resolution data. Furthermore, from these we selected visually a total of $\sim$\,370 point-like, isolated, and uniformly distributed on the OTELO's field sources ($\sim$\,33 objects per \hst\ tile) as a final astrometric reference catalogue. 

The astrometric reference catalogue was cross-matched with a source catalogue, which corresponds to each individual \hst\ $I$-band tiles using \texttt{IRAF\,ccxymatch} task. The astrometric solution was obtained using \texttt{IRAF\,ccmap} third-order polynomial geometry with standard \texttt{TNG} projection. The accuracies of our astrometric solution in standard coordinates $\xi_{I}$ and $\eta_{I}$ are: 0.016 and 0.013 arcsec, respectively. This gives us off-set of 0.021 for the whole set of \hst\ $I$-band tiles -- namely, sub-pixel accuracy in high-resolution \hst\ data with respect to CFHTLS  D3-25 $i'$-band catalogue. The registration process of \hst\ $V$-band images was based on the already aligned $I$-band data. The internal (i.e. between \hst\ $V$- and $I$-band images) astrometric off-set is 0.002 arcsec. 

The final mosaic of scientific and weight (inverse variance) images was obtained with \texttt{SWarp}\footnote{\url{http://www.astromatic.net/software/swarp}} software (\citealt{Bertin2002}). The mosaic was trimmed to the field of \otelodeep\ image with additional $\sim$\,3 arcsec per side in order to provide morphological analysis of the objects which are on the border of the OTELO field of view.

\begin{table}
        \caption[]{Main configuration parameters used in \se\ HDR run.}
        \label{tab:sextractor}
        \begin{tabular}{rll}
                \hline\hline 
                Parameter &  hot & cold \\ 
                \hline
                \texttt{DETECTED\_MINAREA} [px]         & 5     &5 \\
                \texttt{DETECT\_THRESH} [$\sigma$]      & 1.8   &6.0  \\
                \texttt{ANALYSIS\_THRESH} [$\sigma$]    & 1.6   &6.55  \\
                \texttt{FILTER\_NAME}           & \texttt{gauss}$^{(a)}$ & \texttt{tophat}$^{(b)}$  \\
                \texttt{DEBLEND\_NTHRESH} [branch]      & 64    &64 \\
                \texttt{DEBLEND\_MINCONT} [fraction]    & 0.005 & 0.002 \\      
                \texttt{BACK\_SIZE} [px]                & 128   & 256 \\        
                \texttt{BACK\_FILTERSIZE} [px]  & 5     & 9 \\          
                \texttt{BACKPHOTO\_TYPE}        & LOCAL & LOCAL \\      
                \texttt{BACKPHOTO\_THICK} [px]  & 48    & 100 \\        
                \hline
        \end{tabular}
        \tablefoot{$^{(a)}$\texttt{gauss\_4.0\_7x7}; $^{(b)}$\texttt{tophat\_5.0\_5x5}}
\end{table}

\section{Methods}
\label{sec:methods}
\subsection{Parametric classification}
\label{sec:gala_method}
In this work, we used the \GALA\ software\footnote{\url{https://www.nottingham.ac.uk/astronomy/megamorph/}} (\citealt{galapagos2013MNRAS.430..330H}), which, in turn, employs the applications \se\ and \texttt{GALFIT} (\citealt{Galfit2002}; in this case its new, multi-wavelength version \galfitm) to detect and to fit a single-S\'ersic profile to each source, respectively. We chosen this parametric method due to its fully automatic operation: it detects, prepares initial values, runs \galfitm, reads-out the result for each source, and prepares a final catalogue. Furthermore, it handles the issues of neighbouring sources and sky background estimation in an accurate and efficient way \citep{galapagos2013MNRAS.430..330H}.  However, one of the most important characteristic is that \galfitm\ fits simultaneously the selected model to all given bands using a chosen Chebyshev polynomial function. In this work, we used two photometric bands that increase  the number of analysed sources and the accuracy of the parameter fitted to sources (especially with low S/N ratio) as compared with \citet{Griffith2012ApJS..200....9G}. For more, see Section \ref{sec:comparision_with_griffith} (also see simulations in \citealt{galapagos2013MNRAS.430..330H}, their Section 2).

We used the \hstasc\ $I$-band images for source detection in the \se\ dual mode, as implemented in \GALA. Furthermore, following \citet[][]{Rix2004ApJS..152..163R}, so called high-dynamical range (HDR) is used in order to maximise the source detection in two separate \se\ runs with different parameter configurations. The first one (hot run), is optimized to detect faint sources, while the second one (cold run) is optimized to detect bright sources. In Table \ref{tab:sextractor}, we show the values of relevant parameters associated to hot and cold run. Since OTELO goes deeper in magnitude than ACS-GC catalogue, we decided to push the source detection in the faint end, thus, the \se\ parameters were selected via trial and errors. After the hot and cold runs, \GALA\ takes care of matching both catalogues in such a way that hot detections inside a defined ellipse of a cold detection are not included \citep[see Figure 4 from][]{Rix2004ApJS..152..163R}.
We set the following \galfitm\ constraint values \citep[the same as in][]{galapagos2013MNRAS.430..330H}: 
1) position of the object that is to lie within the image cutout (hard-coded into \galfitm);
2) S\'ersic index: $0.2\,<\,n\,<\,8$;
3) effective radius: $0.3\,<\,r_{\rm e, GF}\,<\,400$ [px]; GF stands for the \galfitm\ output;
4) modeled magnitude: $0 < mag_{\rm GF} < 40$. Furthermore $mag_{\rm input} - 5 < mag_{\rm GF} < mag_{\rm input} + 5$, where $mag_{\rm input}$ is the input magnitude \texttt{MAG\_BEST} from \se\ and ${\rm mag}_{\rm GF}$ is the output magnitude  fitted with \galfitm;
5) axis ratio: $0.0001 \leq Q \leq 1$, even if limits $0 \leq Q \leq 1$ are hard-coded in \galfitm\ (based on the recommendation given by \citealt{Haussler2007});
6) position angle: $-180^{\circ}\, < PA < 180^{\circ}$ (hard-coded into \galfitm).

Most of the input parameters in the configuration file were set to the default values, except for the parameters linked with the data used (pixel size, zero-points, exposure time, etc.). During the \galfitm\ fitting we fixed the center position of source to the {\tt X\_IMAGE} and {\tt Y\_IMAGE} $I$-band position, while linear Chebyshev polynomial function is used to fit magnitudes, effective radius $r_{\rm e}$, and S\'ersic index \sersic\ over both \hstasc\ bands used.
In Section \ref{sec:morphological_analysis}, the results obtained from the parametric method described above are compared with the LT and ET classification based on the best SED-fitting templates provided by the \texttt{LePhare} code (Section \ref{sec:otelo_catalogue}).

\subsection{\se\ derived parameters}
\label{sec:sex_method}
A separate \se\ run was executed to fully explore the high-resolution \hstasc\ images. In this run \se\ was fed with the same setup parameters (Table \ref{tab:sextractor}) and carried out in the same dual-HDR mode as in \GALA. This assure the exact correspondence of sources detected in \GALA\ and this separated run. As the output we obtained, for each band used in this work, observed magnitudes (\texttt{MAG\_AUTO}), flux radii (\texttt{FLUX\_RADIUS}) corresponding to radii containing 20, 30, 50, 80, and 90\% of the flux (used to derive the concentration index). 
We decided to perform this run because the output \se\ catalogue from \GALA\ provide the parameter only for the detection band.

\begin{figure}[t!]
        \centering
        \includegraphics[width=\linewidth]{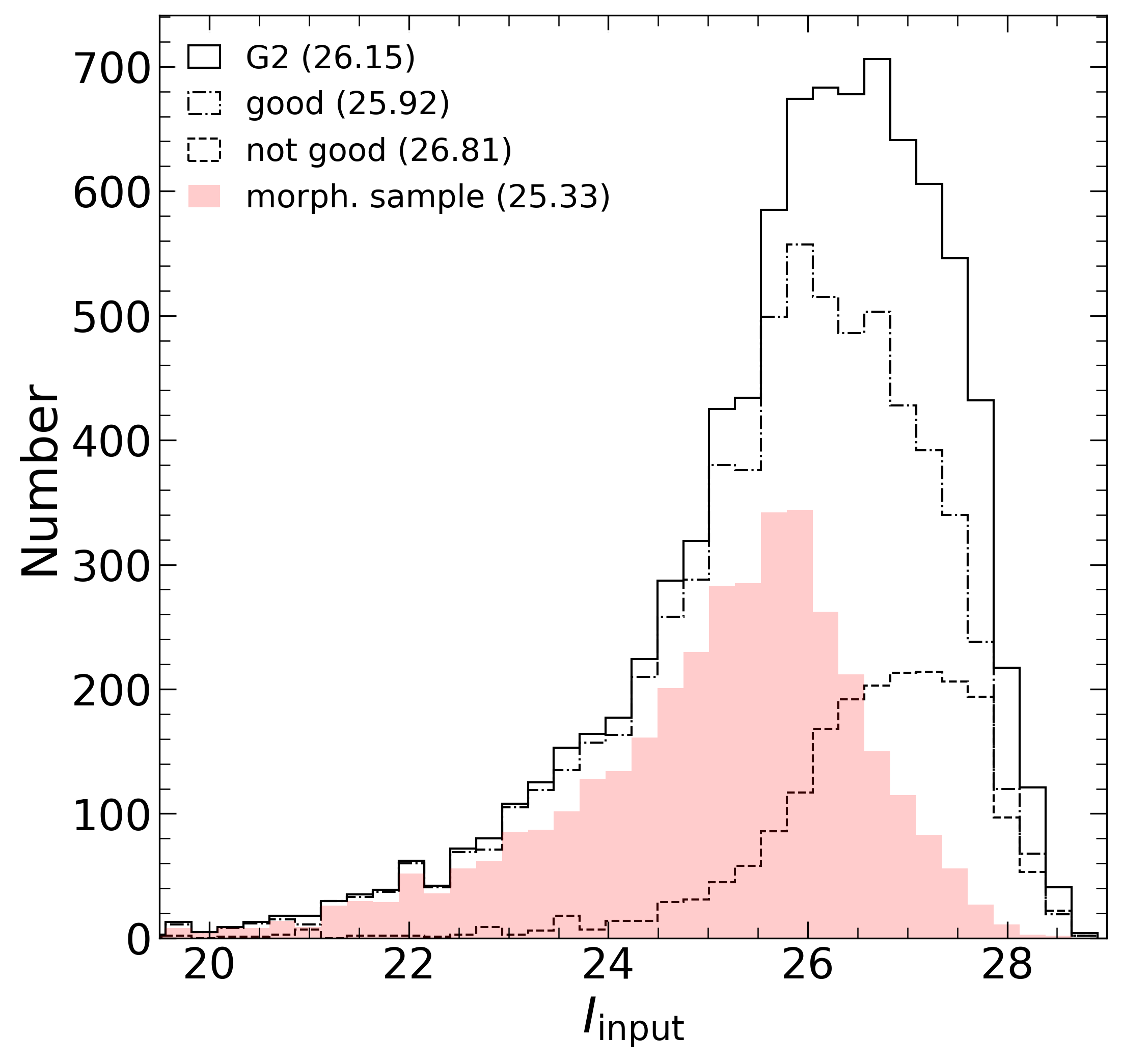}
        \caption[]{Magnitude completeness. Continuous and dot-dash histograms in black represent the G2 and the 'good' samples, respectively. Dashed histogram in black shows the distribution of the sources removed from the G2 sample after the cleaning-out process (see Sec. \ref{sec:sample_selection_galapagos_cat}). Red histogram shows the distribution of the \morphsample\ (see Sec. \ref{sec:sample_selection_gala_vs_otelo_match}). Magnitudes indicated in the legend are 50\% completeness magnitudes for each sample.} 
        \label{fig:mag_completness}
\end{figure}

\begin{figure}[t!]
        \centering
        \includegraphics[width=0.45\linewidth]{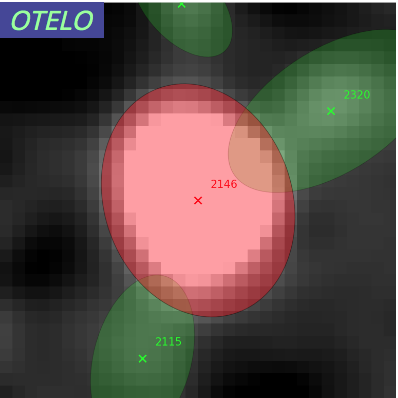}
        \includegraphics[width=0.45\linewidth]{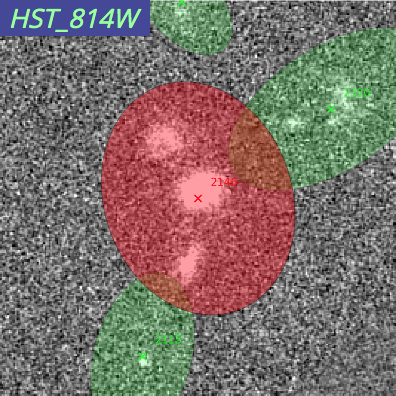}
        \caption[]{Example of a multiple source with central multiple main objects (these images are available in our web-based graphic user interface). {\it Left}: Low-resolution \otelodeep\ image. {\it Right}: \hst\ $I$-band high-resolution image. Ellipses in red (object in question) and green (neighbouring sources) show Kron ellipses from \otelodeep\ image for individual OTELO sources. The red ellipse shows the Kron ellipse corresponding to the source {\tt id:2146}, inside which two additional sources (or well defined parts of a single galaxy) are visible. This particular source has three matched sources from \hst\ image.
}
        \label{fig:2146_match_comp}
\end{figure}

\section{Sample selection}
\label{sec:sample_selection}

As established above, the goal of this work is to assign a morphological classification to the largest possible number of the OTELO survey sources detected in the \otelodeep\ image. Since the \otelodeep\ has lower spatial resolution than \hstasc\ data used to provide the morphological catalogue, a careful match between both is needed. In the first part of this section, we describe the 'cleaning' process of the morphological \GALA\ catalogue obtained from analysis of the \hstasc\ data, while in the second part, we explain the match and further selection of the galaxies, which are analysed in subsequent sections.

\subsection{\GALA\ catalogue}
\label{sec:sample_selection_galapagos_cat}

\begin{figure}[t!]
        \centering
        \includegraphics[width=\linewidth]{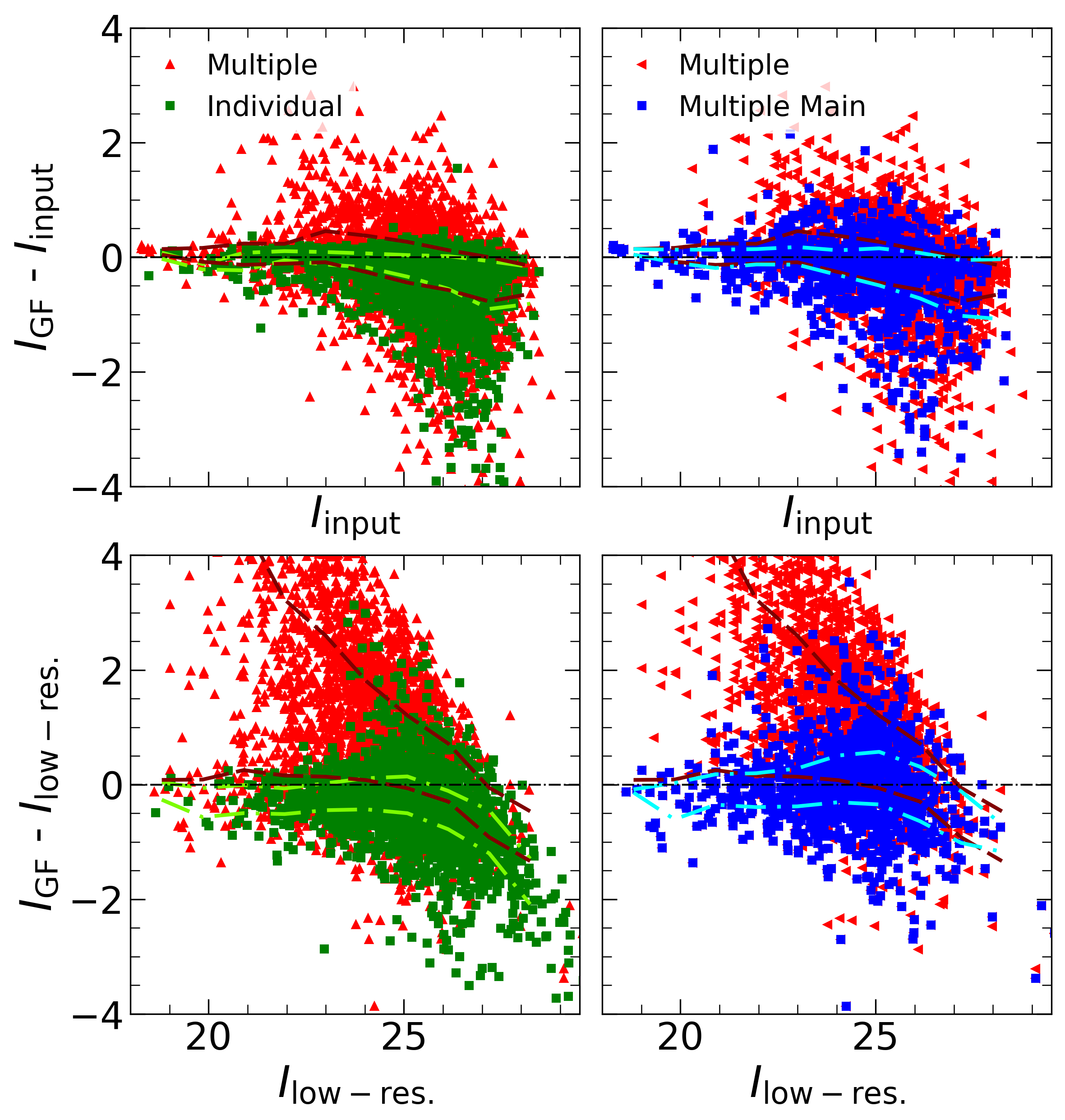}
        \caption[]{Magnitude comparison between low- and high-resolution data. Red up- and left-triangles show multiple sources, green squares show {individual} sources, while blue squares shows {multiple main} sources. Top row: comparison of \hstasc\ input and \galfitm\ output $I$-band magnitudes. Bottom row: the same comparison as in top row, but instead of input high-resolution photometry we plot low-resolution \hstasc-F814W photometry from OTELO catalogue. Lines represent the 25th and 75th percentile of the colour distribution per magnitude bin: light green dot-dashed - individual; dark red dashed - multiples; light blue dot-dashed - multiple main.  See text for details on the samples selection (Section \ref{sec:sample_selection}).}
        \label{fig:mag_comparison}
\end{figure}

\begin{figure}[t!]
        \centering
        \includegraphics[width=\linewidth]{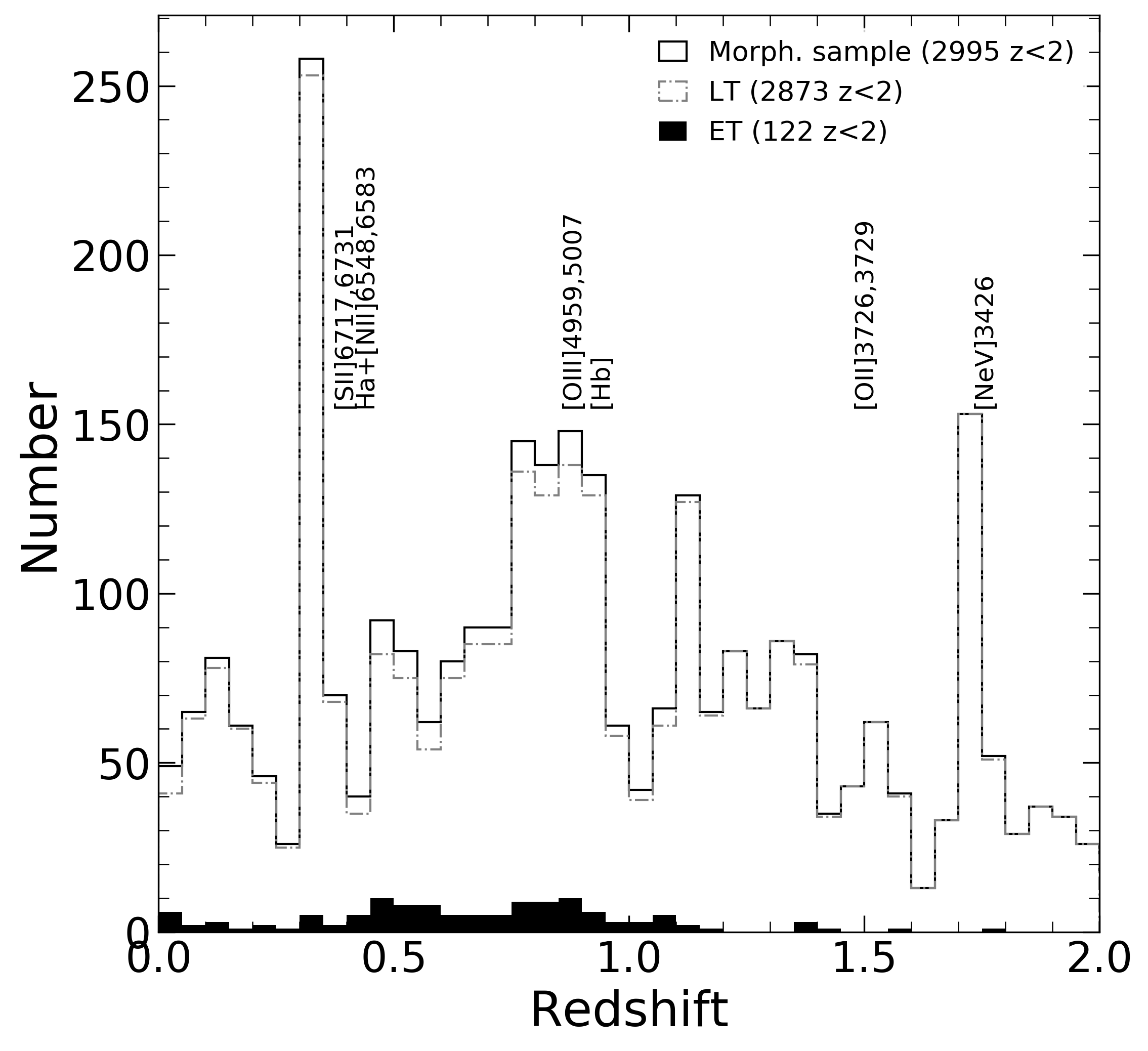}
        \caption[]{Redshift distribution of the selected \morphsample\ (see text for details). Peaks in the distribution corresponds to specific emission lines, as seen in OTELO spectra range up to \z=2. Solid black line shows all sources in \morphsample, gray dot-dashed line shows LT galaxies, while black filled histogram shows ET galaxies.}
        \label{fig:hist_z_dist_both}
\end{figure}

Using the \se\ dual HDR mode, we obtained 10\,591 raw detections from the \hst-I band. A total of 1778 (17\%) raw detections were visually determined to be not real objects. Such high percentage is due to the fact that: (i) the final mosaic of \hstasc\ images (Sec. \ref{sec:astrometry}) still shows imperfections and artifacts, especially on the borders of individual images; (ii) the \se\ parameters were adjusted (via trial and error) to extract the maximum number of real sources, which on the other hand, increases detection of spurious sources. If our visual inspection omitted some of these bad detections, these were removed in the catalogue cleaning process, which is described in what follows. The list of coordinates of these artifacts was passed to \GALA\ in order to omit them in the fitting process.

A total of 8812 (out of 8813) sources were successfully fitted by \GALA\, that is, \galfitm\ does not crush, or does not exceed the time-limit for fitting process. We refer to this successfully fitted sources as G2 sample. The high rate of the successful fits is attributed to the visual inspection of the raw detections. 
The G2 sample, however, still contains sources for which one or more of the fitted parameters hit the constraint value  (listed in Sec. \ref{sec:gala_method}) in one or both \hstasc\ filters, thus, these results are not necessarily meaningful. Using criteria from \citet[their Sec. 4.2]{galapagos2013MNRAS.430..330H} we identify sources characterised by these not necessarily meaningful results and we do not include them in the further analysis. Here, we list all the used criteria applied to both bands:
1) $0\,<\,mag_{\rm GF}\,<\,40$;
2) $0.205 < n <  7.95 $;
3) $0.301 < r_{\rm e} <  399 $;
4) $0.001 <  Q \leq 1$;
5) {\tt FLAG\_GALFIT}$\, =\, 2$.
We omit the criterion (vi) from \citet{galapagos2013MNRAS.430..330H} due to our own star selection described in \citet{OteloI}. Applying those criteria we obtained a sample of 6780 sources with meaningful, or 'good' results (i.e. 77\% of G2 sample, similarly as what is reported in \citealt{galapagos2013MNRAS.430..330H}).  The cleaning process of the morphological catalogue, which involves removing results that are not meaningful for our study guarantees quality results in the following analysis. As shown in Figure \ref{fig:mag_completness}, this process removes the faint end of the G2 sample. The 50\% magnitude completeness drop from $\sim\,26.2$ to $\sim\,25.9$ [AB] for the 'good' sample. Going forward, we use the G2 'good' sample in subsequent sample selections.

\subsection{Matching the \GALA\ and OTELO catalogues}
\label{sec:sample_selection_gala_vs_otelo_match}
We want to stress the importance of a reliable match between the results of the analysis of high-resolution \hstasc\ images using \GALA\ and data from the OTELO catalogue, which are based on the low-resolution ground-based observations (photometry, photometric redshift, SED templates, ET-LT classification, and stellar masses, to name the relevant parameters used in this work). If we want to use any information derived from low-resolution data, we need to assure an exact correspondence between the sources coming from the high-resolution data. Due to the difference in spatial resolution, each source of the OTELO catalogue would admit one or multiple matches to the G2 sample. Hence, we checked the occurrence of one or more sources from G2 'good' sample inside of the \otelodeep\ detection ellipse (see example of multiple match in Figure \ref{fig:2146_match_comp}; ellipses were calculated using \se\ parameters: \texttt{A\_IMAGE $\times$ KRON\_RADIUS} in arcsec). In this match, we got 5338 out of 6780 sources from the G2 'good' sample. At this point, we removed objects which were classified as preliminary star candidates in the OTELO field \citep[see Sect. 6.1 in][]{OteloI}, which gives the total number of 5263 sources. Among these, there are 2780 individual  (i.e. there is one source from G2 sample inside OTELO ellipse) matches between G2 sample and OTELO catalogue, while remaining 2483 sources are matched to 1295 OTELO sources, that is, there is more than one source inside OTELO ellipse (e.g. Fig \ref{fig:2146_match_comp}), and we refer to these as multiple matches. For each multiple match, we selected one source that is the closest to the OTELO catalogued position (usually the brightest counterpart), and we refer to these as multiple main. A total of 1108 multiple main sources were selected. A good example of multiple main is the central source, shown in the right panel of Fig. \ref{fig:2146_match_comp}. 
This is a orthodox way of selecting a fairly clean sample in order to use data obtained from low-resolution OTELO data-products such as \photz, ET-LT classification, or stellar masses. 

The visual inspection of the removed sources during the matching and cleaning out process reveals that these are, in many cases, point-like or very compact sources (possibly QSO), followed by those cases where interaction or merging sources are visible. There are also instances of sources which were excessively deblended by {\se} (i.e. well-resolved galaxies with more than one visible part).

In Figure \ref{fig:mag_comparison}, we show the difference of the output \galfitm\ model magnitude (\magiGF) and the input high-resolution $I$-band \hst\ photometry (\magiGF\, $-\, I_{\rm input}$), as well as the low-resolution photometric data of the PSF-matched \hstasc\ $I$-band from OTELO catalogue (\magiGF\, $-\, I_{\rm low-res.}$). This Figure illustrates the importance of the match and selection process described above.
This figure shows individual (green squares) together with multiple matches (red up-triangles) in the left column, while multiple (red left-triangles) and selected multiple main (blue squares) sources are on the right side. Considering the upper-left panel, we can clearly see that for individual sources, \galfitm\ returns expected values. The dispersion (upper-left panel, green squares) of the magnitudes towards brighter-output \galfitm\ magnitudes is expected due to the integration of the single-S\'ersic model to infinity \citep[e.g.]{Haussler2007,galapagos2013MNRAS.430..330H}. While multiple main (upper-right panel, blue squares) behave in a similar way, their matched group members show larger dispersion. This is even more evident in the case of the low-resolution photometry (bottom row). Thus, even if, on one hand, we introduce bias in our sample selection, on the other hand, we guarantee the correspondence of the parameters from low-resolution data-base to the counterparts in high-resolution data.

On the basis of the source matching scheme described above, we focus on the meaningful results of the sources labeled as individual (2780) and multiple main sources (1108). 
{This gives 3888 sources from which 3658 have any \photz\ solution from the OTELO catalogue while 2995 have \photz\ in the range $0\, \leq\, $\, \photz\, $\, \leq 2$ (hereafter \morphsample)}.  Figure \ref{fig:hist_z_dist_both} shows the redshift distribution of the \morphsample, with LT (2873) and ET (122) sources up to \photz\, $=\, 2$. The peaks of distribution correspond to redshifts at which the OTELO survey sees specific emission lines, for example, \photz$\, \sim\, 0.35$, 0.8, and 1.75 for H$\alpha$, [\ion{O}{III}] and [\ion{Ne}{VI}], respectively. We are aware of the possible misclassification of LT and ET using SED templates. However, \cite{JAD} showed that less than 2\% are misclassified using dense neural networks for the subset of the data used in this work (see their Sect. III.1.4). Thus, for the aims of this work, we consider the OTELO's classification as correct.

This restrictive selection process reduces dramatically the size of the final \morphsample\ (see Figure \ref{fig:mag_completness}). However, in order to use parameters derived in previous works (PSF photometry, \photz, templates classifications associated with \photz, stellar masses, and other OTELO-data products), this selection process is the most reasonable and responsible way to minimise possible biases due to the differences in the data used in the parameter estimation.


\begin{figure}[t!]
        \centering
        \includegraphics[width=0.97\linewidth]{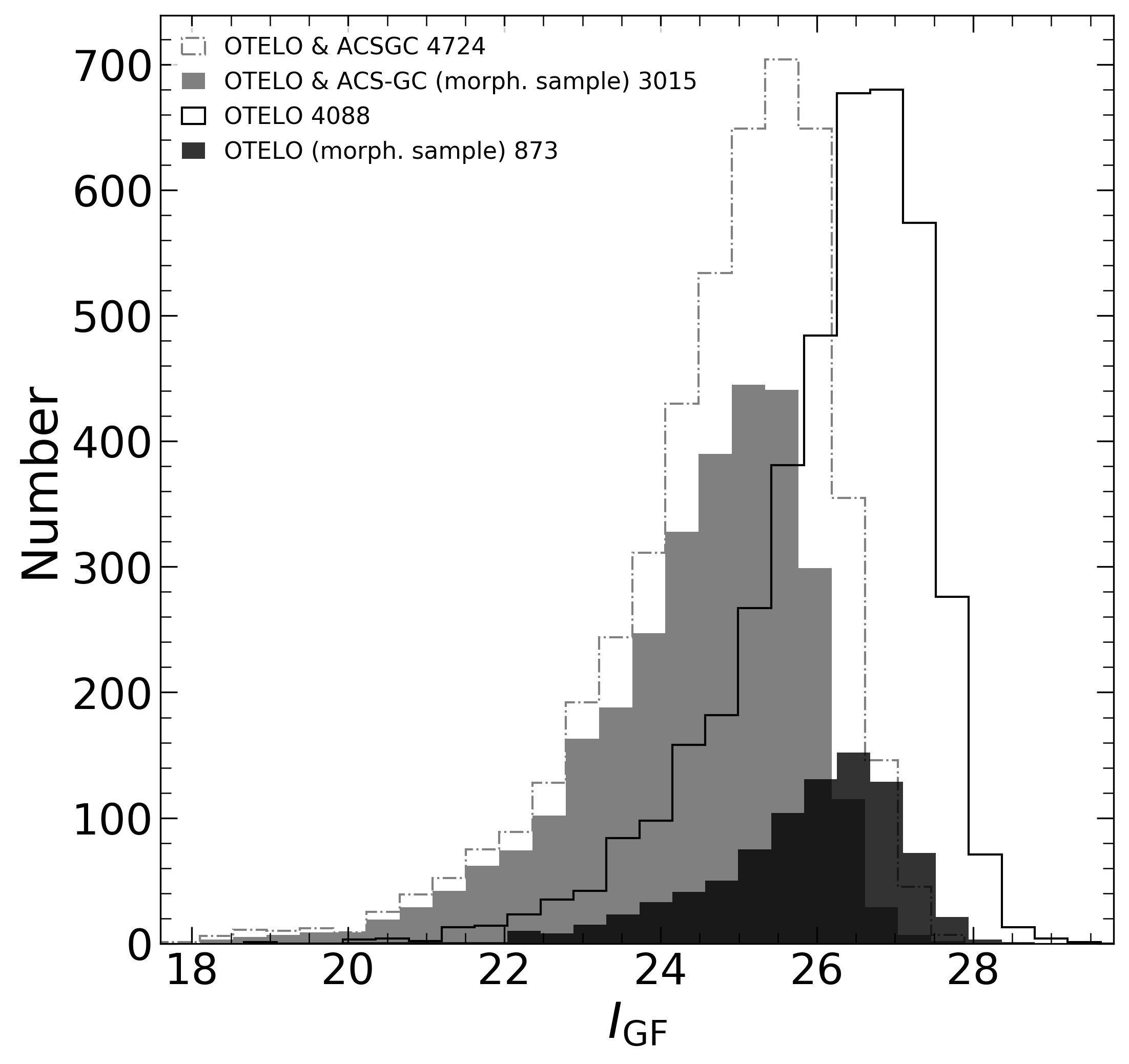}
        \caption[]{{Magnitude comparison with \citet{Griffith2012ApJS..200....9G}. Step histogram in gray (dot-dashed line) show common sample, while step histogram (continuous line) show remaining OTELO sources with no matched counterpart from ACS-GC (i.e. only found in OTELO). Filled gray and black histograms correspond to the \morphsample\ (see Sect. \ref{sec:sample_selection_gala_vs_otelo_match}) from common sample and without counterpart from ACS-GC (i.e. only found in OTELO), respectively. Numbers in the legend indicate the number of sources in the particular sub-sample.}}
        \label{fig:hist_griffith_otelo}
\end{figure}

\subsection{Comparison with ACS-GC catalogue.}
\label{sec:comparision_with_griffith}

\begin{figure*}[t!]
        \centering
        \includegraphics[width=0.97\linewidth]{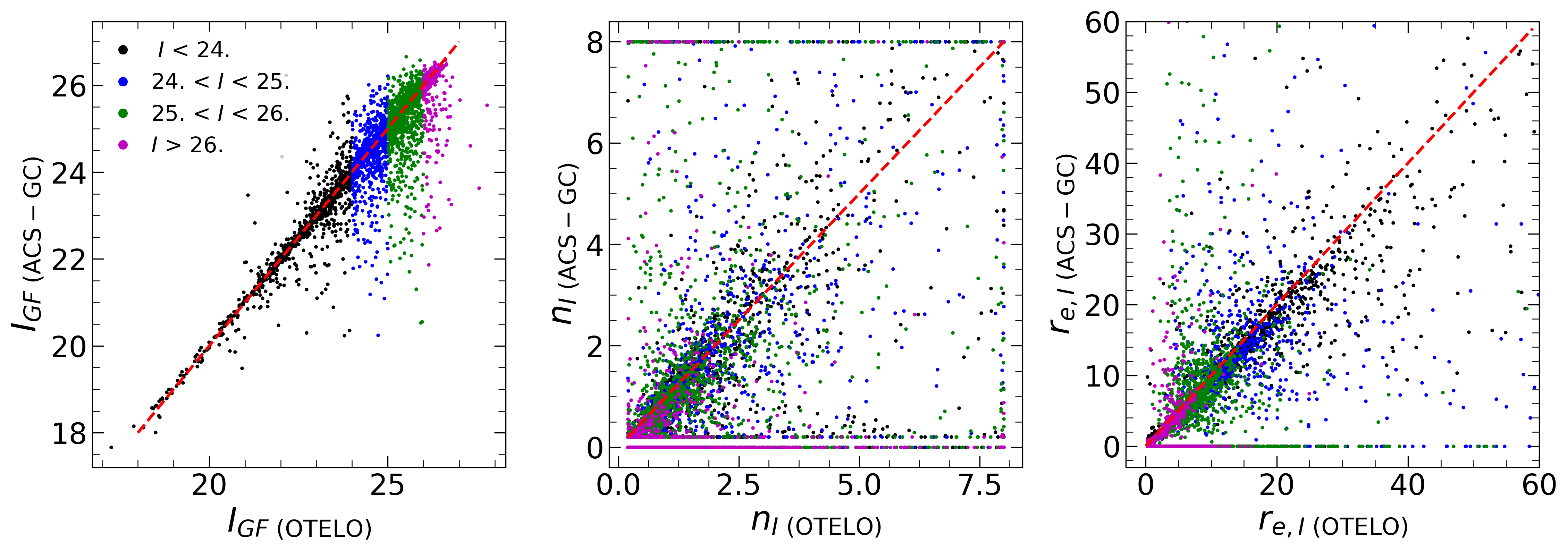}
        \caption[]{{Comparison of the results from this work with those obtained by \citet{Griffith2012ApJS..200....9G} for the common sample. From left to right: model-based $I$-band magnitude $I_{\rm GF}$, S\'ersic index \sersicI, and effective radius \ReGFI. Sources corresponding to each magnitude bin are colour-coded, as shown in the legend on the left panel. Red dashed lines represent 1:1 relation.}}
        \label{fig:mag_n_re_griffith_otelo}
\end{figure*}

\begin{figure*}[t!]
        \centering
        \includegraphics[width=0.97\linewidth]{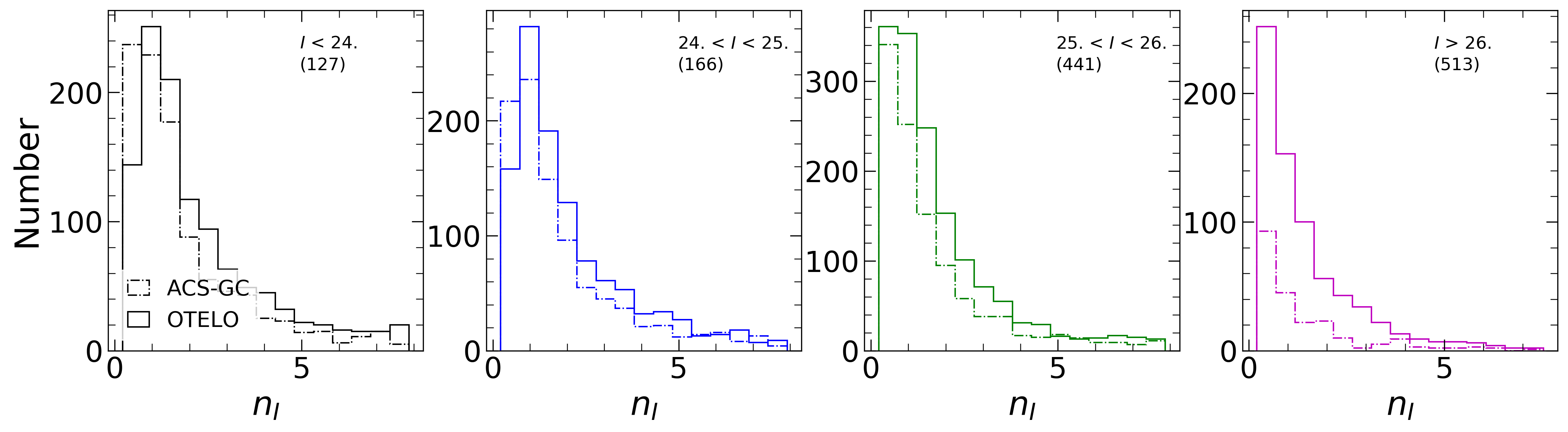}
        \includegraphics[width=0.97\linewidth]{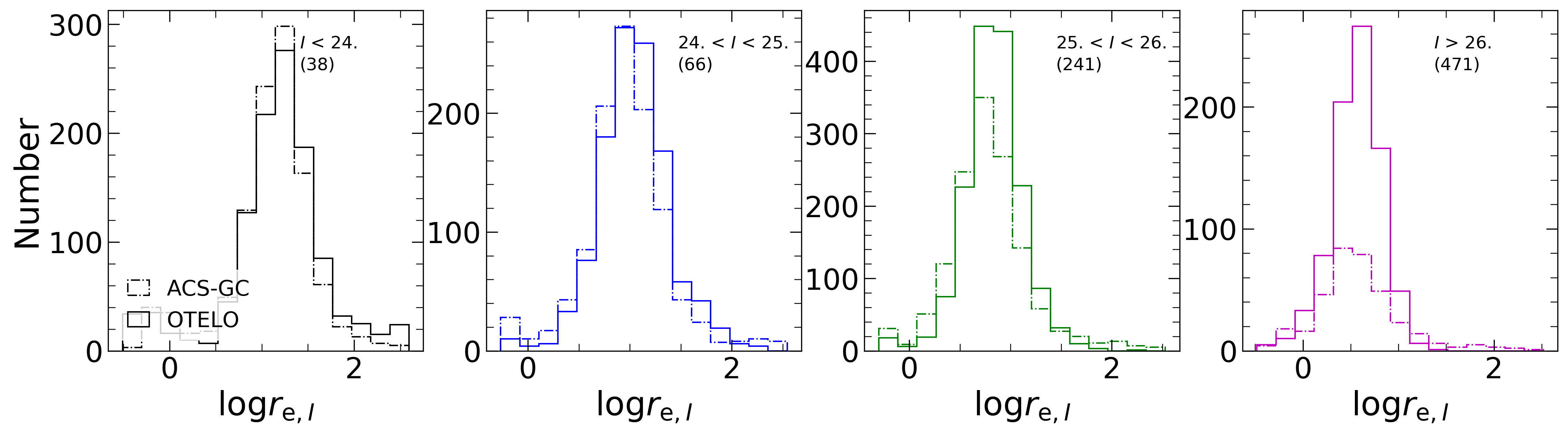}
        \caption[]{Comparison of the results from this work with those obtained by \citet{Griffith2012ApJS..200....9G}. Top row contains a comparison of the S\'ersic index \sersicI\ distributions, while the bottom row represents the logarithm of effective radius \ReGFI. Each panel show a comparison for each magnitude bin indicated in the top-right corner. Numbers between brackets show the number of sources recovered in this work.}
        \label{fig:hist_n_re_griffith_otelo}
\end{figure*}

In this work, we used the same data as previously analysed by \citet[][ACS-GC catalogue]{Griffith2012ApJS..200....9G}, thus, we want to provide a quantitative comparison of the number of successfully analysed sources. The ACS-GC catalogue is based on the standard {\tt GALFIT} version, which treats all bands separately. On the other hand, the multi-wavelength version \galfitm,\ implemented in \GALA\, and used in this work, {allows us to  simultaneously fit the data in all given bands using a linear Chebyshev polynomial.
This multi-wavelength version of {\tt GALFIT} has been shown to be more robust for the low-S/N bands \citep{galapagos2013MNRAS.430..330H}. For the purpose of this comparison, we used a common sample composed of the matched sources from ACS-GC and our G2 sample (with 8812 objects, see Sect. \ref{sec:sample_selection_galapagos_cat} for G2 sample definition). A total of 4724 sources were matched, and these constitute the common sample. The remaining 4088 sources from our G2 sample have no counterpart in ACS-GC (OTELO-only sample). The model-based $I$-band magnitude for both, the common sample, and the OTELO-only sample are shown in Figure \ref{fig:hist_griffith_otelo}. Since the OTELO-only sample may still contain results that are not meaningful for our study, we additionally show the magnitude distribution for each sub-sample which corresponds to the \morphsample\ defined in Section \ref{sec:sample_selection_gala_vs_otelo_match}. From these sub-samples, a total of 3015 sources were found in common (median $I$-band mag. $=$\, 24.4), while the remaining 873 were detected only in the OTELO catalogue (median $I$-band mag. $=$\, 26). We recovered a significant number of robust sources not presented in previous works, which fall into the faint end of the magnitude distribution ($\sim\,$ 1.5 magnitude fainter).}

{Regarding the common sample (4724 sources), in Figure \ref{fig:mag_n_re_griffith_otelo} we show comparison of model-based $I$-band results for magnitudes $I_{\rm GF}$, S\'ersic index \sersicI, and effective radius \ReGFI. In order to make our comparison sensitive to S/N, we divided the common sample into four magnitude bins ($I_{\rm GF} \leq$ 24, 24 < $I_{\rm GF} \leq$ 25, 25 < $I_{\rm GF} \leq$ 26, and $I_{\rm GF}$ > 26; colour-coded in all panels). The left panel shows the magnitude comparison and it is clear that both catalogues are in agreement. The middle and right panels represent the S\'ersic index, \sersicI, and effective radius, \ReGFI\ for a direct comparison of both catalogues, ACS-GC and OTELO. In the middle panel, where \sersicI\ is represented, we can notice that part of ACS-GC results is accumulated in three discrete values, that is, at \sersicI\, $=$\, 0, around 0.2, and 8. These are the results for which {\tt GALFIT} run into constraint values, and should not be considered as valid. 
While this is clearly visible for ACS-GC, much fewer sources are found at these constraint values among the OTELO results. The same is true for the effective radius, \ReGF,\ shown on the right panel, although less pronounced, as compared to the S\'ersic index. 

In Figure \ref{fig:hist_n_re_griffith_otelo}, we show histograms for \sersicI\ and \ReGFI\ per magnitude bins. In these histograms, we include sources from the common sample after applying the criteria 2 and 3 listed in Section \ref{sec:sample_selection_galapagos_cat} (i.e. $0.205\, <\,$\sersicI$\,<\, 7.95$, and $0.301\,<\, $\ReGFI\,$<\, 399$, thus removing sources with results that hit constraint values) for both catalogues. We indicate the difference of the sources in OTELO and ACS-GC for each parameter and bin (numbers between brackets) after applying the aforementioned criteria. As can be seen, for both parameters and all magnitude bins, OTELO has substantially more sources with meaningful results. The greatest difference is observed in the faintest bin, thus confirming that the multi-wavelength version of {\tt GALFIT} used in this work is more robust, especially in the low S/N regime. Similar results are obtained for the remaining parameters, as well as for the $V$-band.
}

\section{Morphology analysis}
\label{sec:morphological_analysis}

Among the data returned by \GALA\ there are model-based parameters like magnitudes (\magvGF, \magiGF), S\'ersic indices (\sersic), and effective radii (\ReGF, containing 50\% of the model flux) for each input filter, that is, \hstasc\ $I$- and $V$-band. The analysis also includes  the parameters obtained from the separate \se\ run, as described in Section \ref{sec:sex_method} (e.g. concentration index), as well as stellar masses, \mstar,\ from \citet{Nadolny}.

\begin{figure}[t!]
        \centering
        \includegraphics[width=\linewidth]{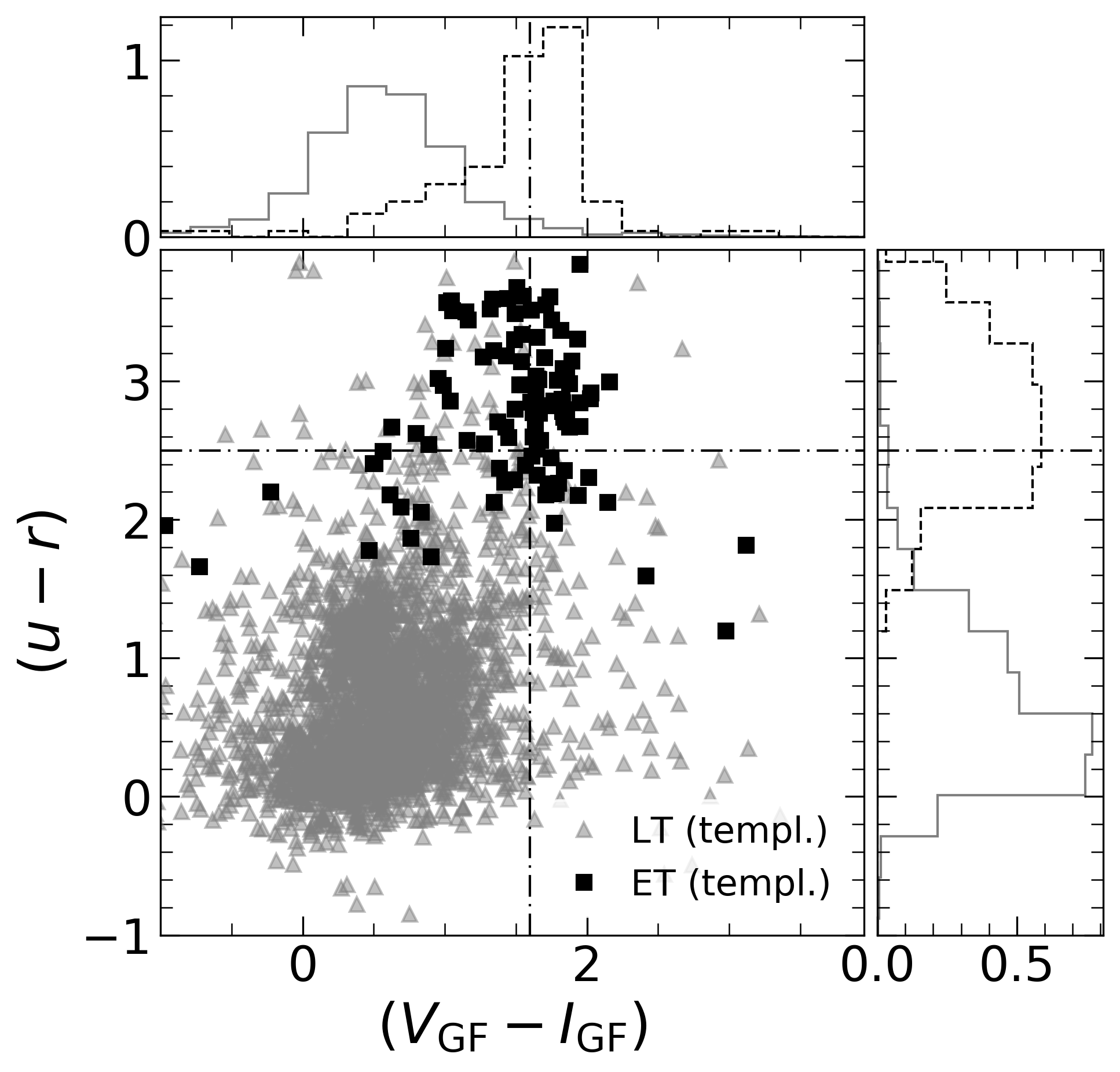}
        \caption[]{Comparison between \colourVIgf\ and \colourUR\ colours. Triangles and solid lines in gray represent LT, squares and dashed lines in black correspond to ET galaxies. The top and right-hand density histograms correspond to \colourVIgf\ and \colourUR, respectively. Dot-dashed lines show the results of linear discriminant analysis of both colours with \colourVIgf$=1.7$ and \colourUR$=2.5$.} 
        \label{fig:colour_comparison}
\end{figure}

Figure \ref{fig:colour_comparison} shows the results of a linear discriminant analysis of model-based \colourVIgf\ and observed \colourUR\ colours performed to find the most accurate ET-LT separation. We found cuts of $1.7$ and $2.5$ to be the most accurate for \colourVIgf\ and \colourUR, respectively. Using these limits, we found ET completeness and contamination of 39\% and 63\% for \colourVIgf, and 62\% and 35\% for \colourUR, respectively. The small spectral separation of the filters used to calculate the \colourVIgf\ colour translates into the relatively poor separation of ET from LT galaxies, with a higher contamination of LT in the expected region for ET. The \colourUR\ colour with larger wavelength separation, gives better results in terms of completeness and contamination. This shows that \colourUR\ is more adequate for ET-LT separation. 
The \colourUR\ colour cut found in this work is higher than that reported by \citet{Strateva2001AJ....122.1861S} and we attribute this with the redshift range of our sample. 

The high completeness ($>\,97$\%) of LT in blue cloud is likely due to the selection process, where we cleared out the \morphsample\ from possible QSO or AGN and mergers or interacting galaxies, as described in Section \ref{sec:sample_selection_galapagos_cat}. In order to carry out a second check, we used the criteria from \citet[][their Equations 1 and 2]{Schawinski2014MNRAS.440..889S} to separate blue cloud, green valley, and red sequence galaxies. We found that $\sim\,8$\% of our LT galaxies are in their red sequence as compared with 7\% in their work (see their Table 1). 

\begin{figure*}[t!]
        \centering
        \includegraphics[width=\textwidth]{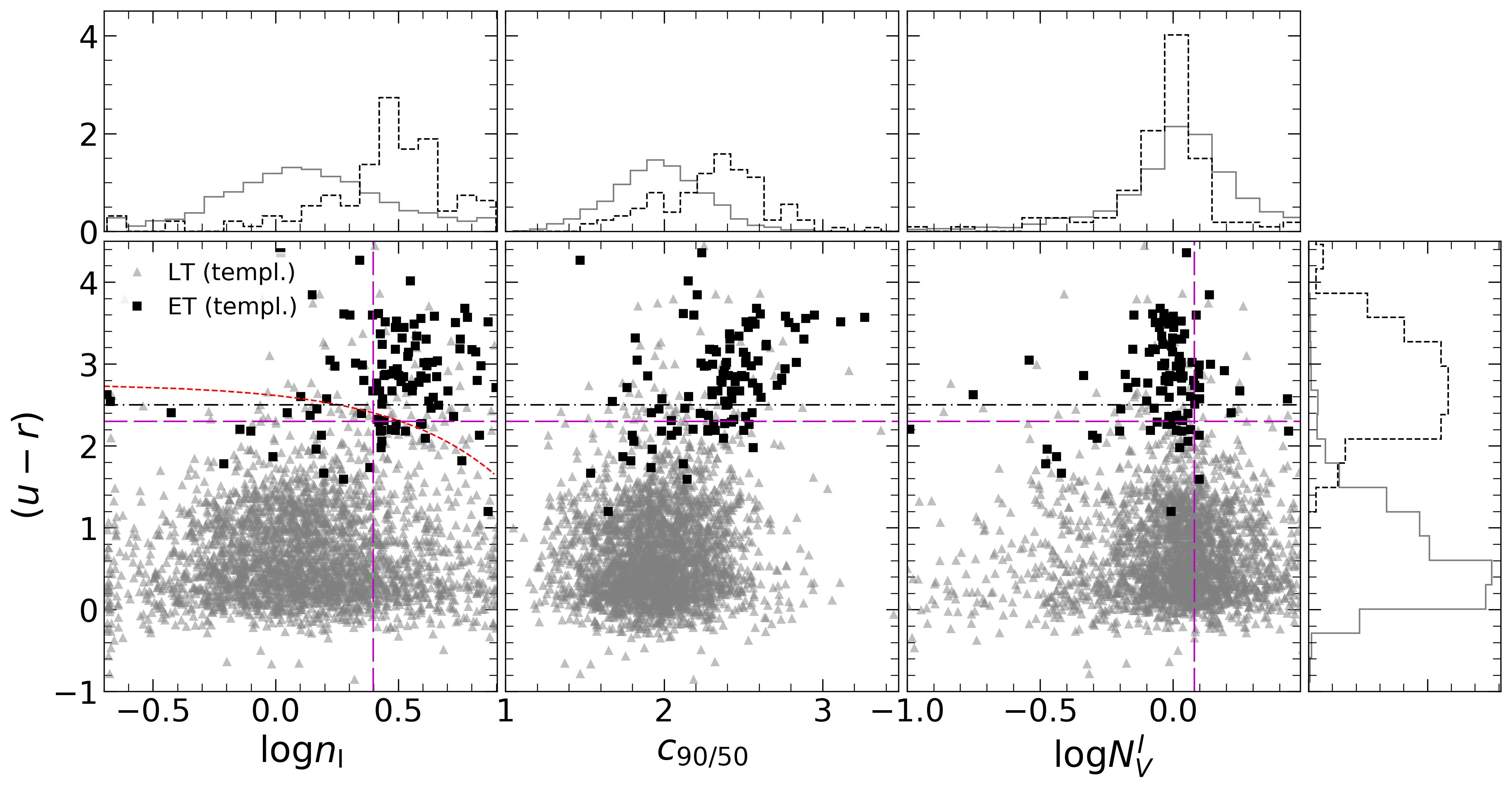}
        \caption[]{Observed \colourUR\ colour as a function of morphological parameters. From left to right: S\'ersic index, \sersicI, concentration index, \concentration,\ and wavelength-dependent ratio of S\'ersic indices \N. Triangles and solid lines in gray (histograms) show LT, squares and dashed lines in black (histograms) show ET galaxies. The top histograms correspond to the respective value, as indicated in x-axis label, while right-hand histogram show the \colourUR\ colour distribution. All histograms represent density distributions. 
        Horizontal dot-dashed line in black shows \colourUR$\,=\,2.5$. Red dashed line show linear discriminant analysis result from \cite{JAD}. Dashed lines in magenta represent limits from \cite{Vika_MegaMorph2015}: vertical cut in \colourUR\,$=\,2.3$, while on the left and right panels horizontal dashed-lines in magenta represent $\log$(\sersicI)\,$=\,0.4$ and $\log$(\N)\,$=\,0.08$, respectively (see Section \ref{sec:discussion}).}
        \label{fig:sersic_concentration_niv}
\end{figure*}

The bottom-left panel of Figure \ref{fig:sersic_concentration_niv} shows the S\'ersic index distribution  as a function of \colourUR\ colour. The overall S\'ersic indices fall in the expected ranges of values for ET and LT galaxies (top-left histogram). The median values of \sersicI\ index are 1.3$\,\pm\,$0.6 and 3.0$\,\pm\,$0.9 for LT and ET, respectively. Uncertainties cited in this section are median absolute deviations. The increase of \sersic\ from LT to ET is expected, since LT are disc-dominated (described with lower S\'ersic index of \sersic$\,\sim\,$1) and ET are bulge-dominated (i.e. \sersic$\,\gtrsim\,$4) galaxies. A very similar trend is observed for \sersicV\ (measured on $V$-band) with median values of 1.1$\,\pm\,$0.6 and 3.0$\,\pm\,$1.2 for LT and ET, respectively. The $V$-band results are not shown for the sake of clarity. 
As can be seen in the same panel of Figure \ref{fig:sersic_concentration_niv}, there are two well-defined regions occupied by red \& high-\sersic\ ET galaxies [\colourUR\,>\,2.3 and $\log\,$\sersic\,>\,0.4] and blue \& low-\sersic\ LT galaxies [\colourUR\,<\,2.3 and $\log\,$\sersic\,<\,0.4]. We found a ET completeness in the red \& high-\sersic\ zone of 59\%, with a contamination of 30\%. The completeness (and contamination) provided by this method according to \citet{Vika_MegaMorph2015} is of 63\% (53\%) for their artificially redshifted sample (see their Table 1). In the same panel of Figure \ref{fig:sersic_concentration_niv}, we show the results of the linear discriminant analysis from \cite{JAD}. Using this discriminant for the \morphsample,\ we found 98\% and 66\% of completeness for LT and ET, respectively.

The concentration index \concentration, defined as ratio of \se\ $I$-band flux radius containing 90\% and 50\% of the flux, is shown in the middle panel of Figure \ref{fig:sersic_concentration_niv}. The median values of $I$-band \concentration\ are 1.9$\,\pm\,$0.2 and 2.4$\,\pm\,$0.2 for LT and ET, respectively. Again, very similar median values are observed for $V$-band \concentration\ with 1.9$\,\pm\,$0.2 and 2.3$\,\pm\,$0.2 for LT and ET. As shown in previous works (e.g. \citealt{Strateva2001AJ....122.1861S}), this parameter does not provide a good separation of ET-LT and it can only yield a crude classification.

The last parameter in  Figure \ref{fig:sersic_concentration_niv} (third panel) shows the the ratio of S\'ersic indices in both \hst\ bands \N\, $=$\,\sersicI/\sersicV, as defined by \citet{Vika_MegaMorph2015}. This parameter, together with a colour term, is shown to be sensitive to the internal structure. Median values of \N\ for LT and ET are 1.1$\,\pm\,$0.4 and 1.0$\,\pm\,$0.2, respectively. This indicates that the ET sub-sample is well defined around \N\,$= 1$. This is an expected result because of the very nature of red ET galaxies, which show less variation of the S\'ersic profile with wavelength (e.g. \citealt{Vulcani_MegaMorph_2014}). Using cuts of \N\,<\,1.2 and \colourUR\,>\,2.3, \citet{Vika_MegaMorph2015} found 70\% of completeness and 50\% of contamination for ET galaxies. Here, we report 60\% of ET completeness with 42\% of LT contamination using the same cuts in \N\ and \colourUR. 

\subsection{Testing the mass-size relation}
\label{sec:morphological_analysis_msr}
\begin{figure*}[t]
        \centering
        \includegraphics[width=\linewidth]{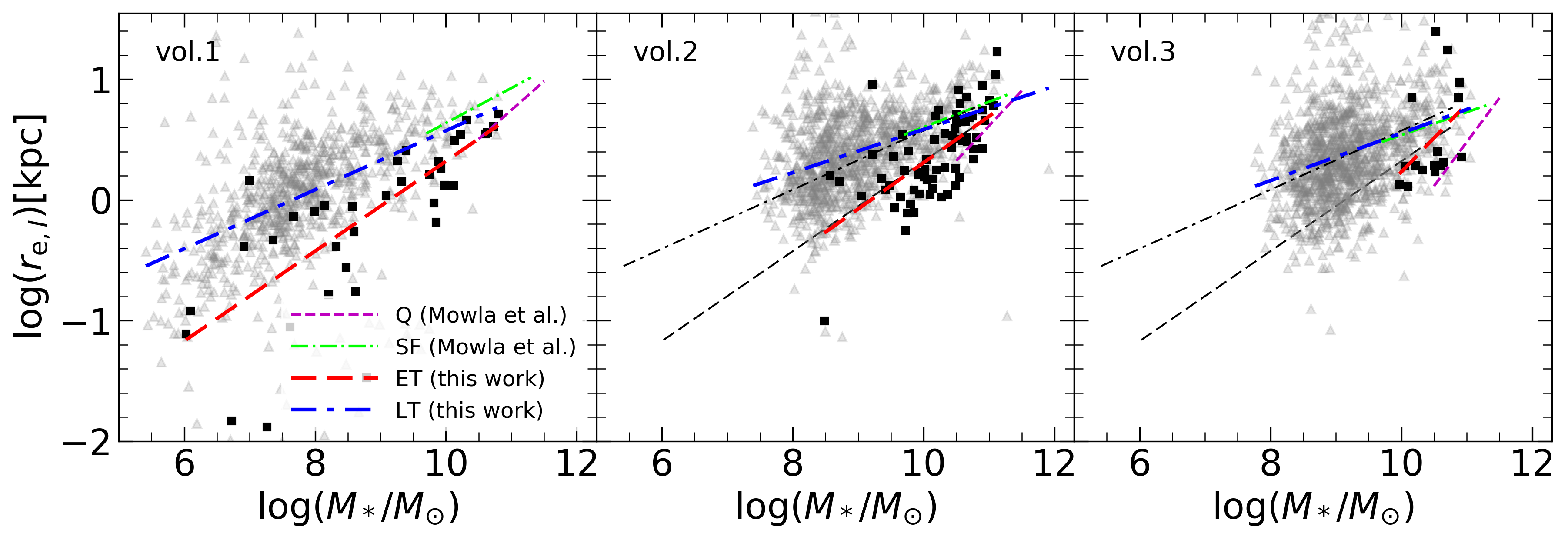}
        \caption[]{ {Mass-size relation. Each panel correspond to selected cosmic volumes: $z < 0.5$, vol2: $0.5 \leq z < 1$, and vol3: $1 \leq z < 2$. Gray triangles and black squares show LT and ET galaxies, respectively. Dot-dashed blue and dashed red lines show power--law fits for LT and ET data, respectively, according to Eq. \ref{eq:fiteq}. The best-fit parameters are given in Table \ref{tab:MSR}. Gray dot-dashed, and dashed lines in the middle and right panels show the fits for LT and ET obtained for the first cosmic volume (vol1). Dot-dashed green and dashed magenta lines represent LT (SF) and ET (Q) MSR from \cite{Mowla2019ApJ...880...57M}.}}
        \label{fig:mass_size_per_volume}
\end{figure*}

Using stellar masses, \mstar,\ and physical sizes estimated using model-based flux radius, \ReGF,\ we can study the mass-size relation (MSR). In order to probe if there are any signs of evolution in the MSR since $z = 2$, we divided our sample into three redshift bins or volumes (vol1: $z < 0.5$, vol2: $0.5 \leq z < 1$, and vol3: $1 \leq z < 2$). {For each bin and for each morphological type,  we fit a single power law of the form:}
\begin{equation}
r_{\rm e} =  a\,({M}_*/{M}_{\rm \odot})^b.
\label{eq:fiteq}
\end{equation}

{\indent The parameters of the best-fit functions for each cosmic volume, as well as median redshifts, are given in Table \ref{tab:MSR}. These functions are represented in Figure \ref{fig:mass_size_per_volume}. As can be seen, the slope of the relation fitted in this process do not vary significantly between different redshift bins for a given morphological type (the black lines in the middle and right panel in Fig. \ref{fig:mass_size_per_volume} show the fits from the first redshift bin). It is noticeable that, on one hand OTELO miss the high-mass end ($> 10^{11}\,$\msun) of the LT and ET galaxies, while on the other hand, there is also a low store of statistics with regard to the ET galaxies in the last redshift bin (resulting in high errors in the best-fit power law; see Table \ref{tab:MSR}). 

The OTELO survey was designed to recover the low-mass end of the field galaxy population. Indeed, in our MSR, we can see that it extends towards lower stellar masses, if compared with previous works. Thus, in all panels of Fig. \ref{fig:mass_size_per_volume} we additionally represent the fits from \citet{Mowla2019ApJ...880...57M} for star-forming (LT) and quiescent (ET) for the same redshift bins. Their separation onto star-forming and quiescent galaxies is based on $U$, $V$, and $J$ rest-frame bands. For the purpose of the comparison, presented in what follows, this selection is compatible with our ET-LT separation \citep[][]{vanderWel2014ApJ...788...28V}. Our MSR for LT galaxies is consistent with those given by \citet{Mowla2019ApJ...880...57M}, even taking into account the differences in the mass range studied in both works. However, the slope obtained by these authors in the case of ET galaxies is stepper than the fitted in this work for all redshift bins, with the largest difference in the last volume $1 \leq z < 2$ (0.73 and 0.56, in \citeauthor{Mowla2019ApJ...880...57M} and in this work, respectively). This difference mainly obeys to the fact that the galaxies studied in \citet{Mowla2019ApJ...880...57M} are massive \logmass$\, > 11.3$, complemented with galaxies from \citet[][3D-HST+CANDELS]{vanderWel2014ApJ...788...28V}. Precisely, this addendum contains a great part of the ET galaxies of the resulting sample in the last redshift bin studied in their work. Since both works, \citet{vanderWel2014ApJ...788...28V}, and \citet{Mowla2019ApJ...880...57M} are consistent, we decided to only keep  the fits from the latter.} 

It is worthwhile noting that due to the limited OTELO's field of view ($\sim$\,56 arcmin$^2$), the co-moving volume surveyed is relatively small if compared with broadly known extragalactic surveys (e.g. SDSS, COSMOS, CANDELS). This fact may introduce a bias towards the detection of sources with lower masses, thus losing the high-mass end. For instance, if the results of the zCOSMOS survey are scaled to the sky area explored by OTELO, and using the same selection criteria given by \citet[][i.e. $0.1 < z < 1.1$, \logmass > 11 and \otelodeep < 24]{LsJ_COSMOS_ET_2012A&A...548A...7L}, we should find 11 sources in the census of the OTELO survey. However, following the selection process described in Sect. \ref{sec:sample_selection}, we are left with only four sources. Hence, it is necessary to stress that we lose $\sim\,60\,$\% of the high-mass end  as a result of the limited sky covering of our survey.

\begin{figure}[t]
        \centering
        \includegraphics[width=\linewidth]{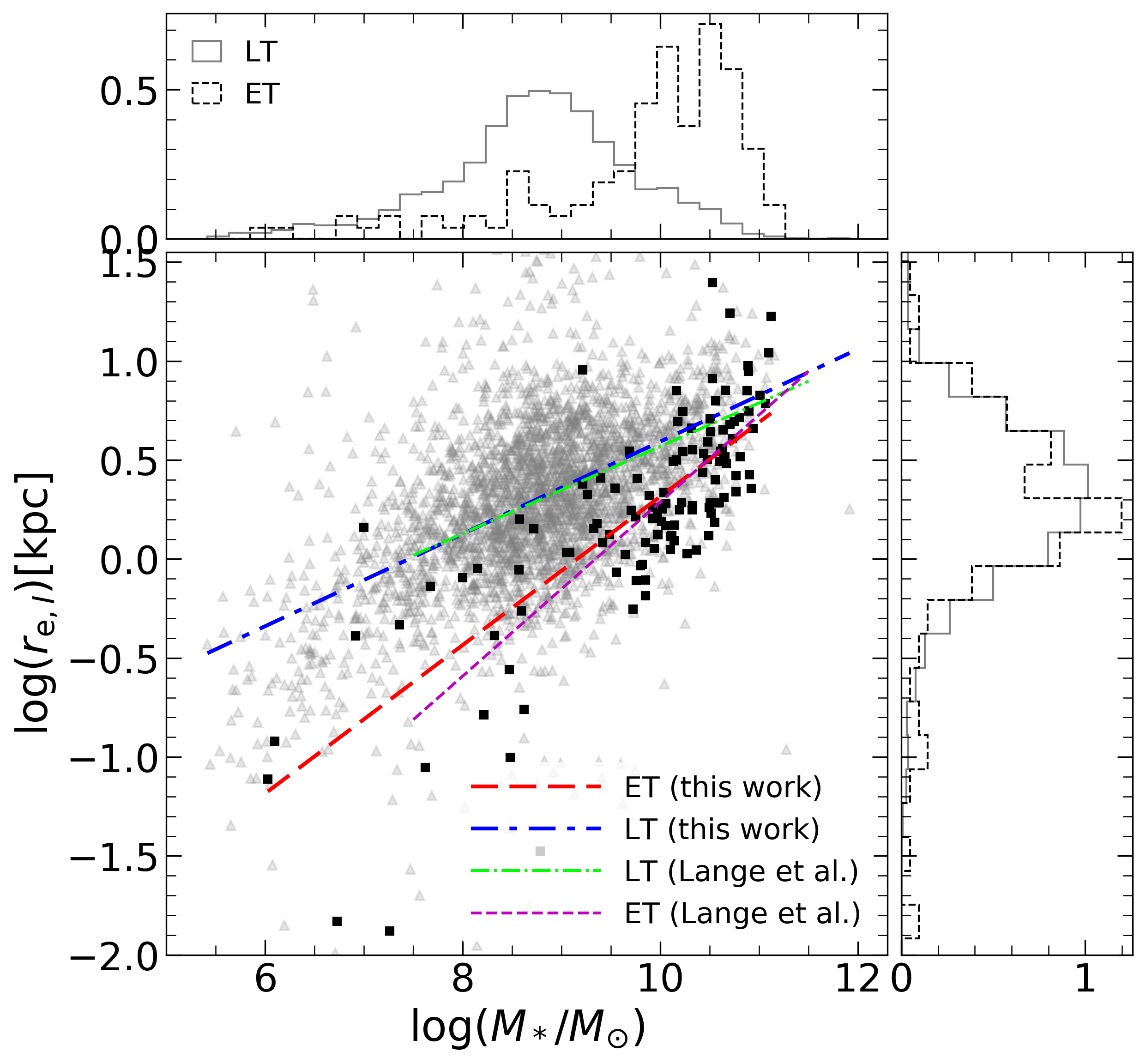}
        \caption[]{Mass-size relation. Gray triangles and black squares show LT and ET galaxies, respectively. Dot-dashed blue and dashed red lines show linear fits for LT and ET data, respectively. Thin dot-dashed green and dashed magenta lines represent LT and ET MSR from \cite{LangeGAMA_Mass_Re_2015}. The top and right panels show density histograms of stellar mass and size, respectively. {The results of the fitted power law parameters are given at the bottom of Table \ref{tab:MSR}.}}
        \label{fig:mass_size}
\end{figure}

{In Figure \ref{fig:mass_size}, we show ET and LT galaxies analysed in this work without separation onto redshift bins. In this figure, we show the power--law fits to the whole ET and LT samples (tabulated in the bottom part of Table \ref{tab:MSR}), as well as results from \citet[][their Table 2 and 3, $i$-band fits]{LangeGAMA_Mass_Re_2015}, where the local galaxy sample was studied ($0.01 < z < 0.1$, \logmass$\, \sim 7.5 - 11$). The MSR for ET and LT populations of OTELO are consistent with the low-redshift relation. In order to test the effects of the selected stellar mass range when comparing our results with previous studies, we limited our sample to the lower limits given in \citet[][]{Shen2003MNRAS.343..978S}. These limits in \logmass\ are 10.1 and 8.8 for ET and LT galaxies, respectively. Both limits correspond to median mass values of ET and LT samples studied in this work (i.e. by removing considerable part of our sample). Even so, using these mass-limited samples, we found a good agreement between our fit and the results of \citet{Shen2003MNRAS.343..978S}. For the sake of clarity, we do not include these results in Figure \ref{fig:mass_size}. In the top panel of the same figure, we show the distribution of stellar masses for ET and LT galaxies. It is noticeable that both populations dominate in different mass regimes, with median values of \logmass$ \sim\, 8.8$ and $10.12$ for LT and ET galaxies, respectively. On the other hand, as can be seen on the right panel of Figure \ref{fig:mass_size}, both galaxy populations occupy the similar range of physical sizes. In Section \ref{sec:discussion}, we present more detailed discussion on possible median size evolution in the $r_{\rm e}$--$z$ space.}

\begin{table}
        \caption[]{Results of the power law fitting of the MSR.}
        \label{tab:MSR}
        \setlength{\tabcolsep}{2.5pt}
        \begin{tabular}{lccccc}
                \hline\hline 
                samples& $<z>$ & $\log\,a$ & $b$ & n & \logmass \\
                 & & & &  & range\\
                \hline

ET vol1 & 0.33 & -3.41  $\pm$ 0.63  & 0.37  $\pm$ 0.07 & 37 & 6.03 -- 10.80 \\
ET vol2 & 0.77 & -3.53  $\pm$ 0.56  & 0.38  $\pm$ 0.06 & 68 & 8.48 -- 11.12 \\
ET vol3 & 1.11 & -5.42  $\pm$ 3.31  & 0.56  $\pm$ 0.32 & 17 & 9.97 -- 10.92  \vspace{.5mm} \\
LT vol1 & 0.34 & -1.87  $\pm$ 0.13  & 0.24  $\pm$ 0.02 & 749 & 5.42 -- 10.78 \\
LT vol2 & 0.79 & -1.21  $\pm$ 0.13  & 0.18  $\pm$ 0.01 & 964 & 7.40 -- 11.91  \\
LT vol3 & 1.39 & -1.41  $\pm$ 0.18  & 0.20   $\pm$ 0.02 & 1160 & 7.77 -- 11.06 \vspace{.5mm} \\
\hline\hline
 ET all & 0.66 & -3.44 $\pm$ 0.32 & 0.38 $\pm$  0.03 & 122 & 6.03 -- 11.12  \\
 LT all & 0.88 & -1.74 $\pm$ 0.06 & 0.23 $\pm$  0.01 & 2873 & 5.42 -- 11.91 \\
                \hline\hline
        \end{tabular}
        \tablefoot{Redshift bins are defined as follows: vol1: $z < 0.5$, vol2: $0.5 \leq z < 1$, and vol3: $1 \leq z < 2$. { Median redshift for each volume and type is given in column 2. For each bin we fit a single power law according to Eq. \ref{eq:fiteq}, with the number of galaxies and stellar mass range indicated in the two last columns. These are shown in Figure \ref{fig:mass_size_per_volume}.} The last two rows give the results of fit using all the sources in ET and LT samples -- both are shown in Figure \ref{fig:mass_size}.}
\end{table}

\subsection{Errors}
\label{sec:errors}
\begin{figure}[t]
        \centering
        \includegraphics[width=\linewidth]{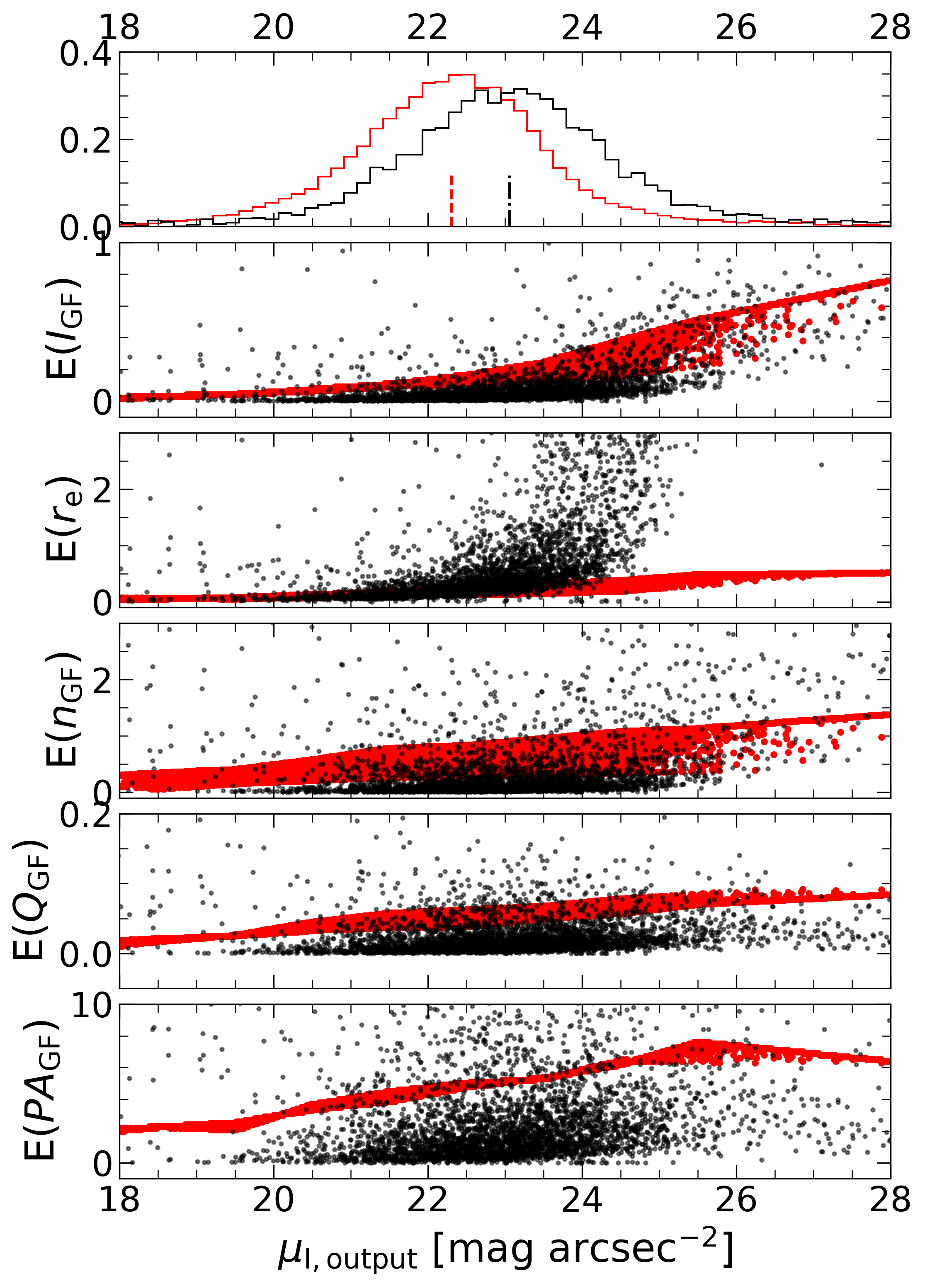}
        \caption[]{Errors of given parameter as a function of surface brightness \muoutput. Black dots show the outcome from this work and the red dots show data for selected (f1=1) galaxies from \citet{Haussler2007}. The top panels show density distribution of \muoutput\ with median values marked as red dashed-lines (22.3) and black dot-dashed lines (23.04).}
        \label{fig:comparison_erros_haussler2007}
\end{figure}

As described in \cite{GALFIT_2010AJ....139.2097P}, the formal uncertainties derived in \galfitm\ are only the lower estimates. After extensive simulations and tests of this software, \citet{Haussler2007} provided estimates of uncertainties based on the comparison of input and output values as a function of surface brightness. In Figure \ref{fig:comparison_erros_haussler2007}, we show the errors from \citet{Haussler2007} and nominal errors from \galfitm\ from this study as a function of the surface brightness \citep[defined as in][]{Haussler2007}, given by:
\begin{equation}
\mu_{\rm output} = {mag + 2.5 \log[2(b/a)\,\pi\,r_{\rm e}^{2}]},
\end{equation}
where $mag$ is the $I$-band magnitude, $b/a$ the axis ratio, and r$_{\rm e}$ the half-light radius in arcseconds. The \muoutput\ is calculated using the output \galfitm\ parameters. We note that even the nominal error values are the lower estimates, as it is clearly seen that these behave as expected -- the errors increase with surface brightness for all parameters (even if this is not so evident for \sersic, \textit{Q,} and \textit{PA}). This is also seen if the errors are represented as a function of $I_{\rm input}$ (however, we do not show this relation here). We report the nominal values of the errors with the indication that these are lower-bound estimates (at least for bright objects) since the total error budget would require performing simulations as in \citet{Haussler2007}, which is beyond the scope of this work.

\section{Discussion}
\label{sec:discussion}
\begin{figure}[t]
        \centering
        \includegraphics[width=\linewidth]{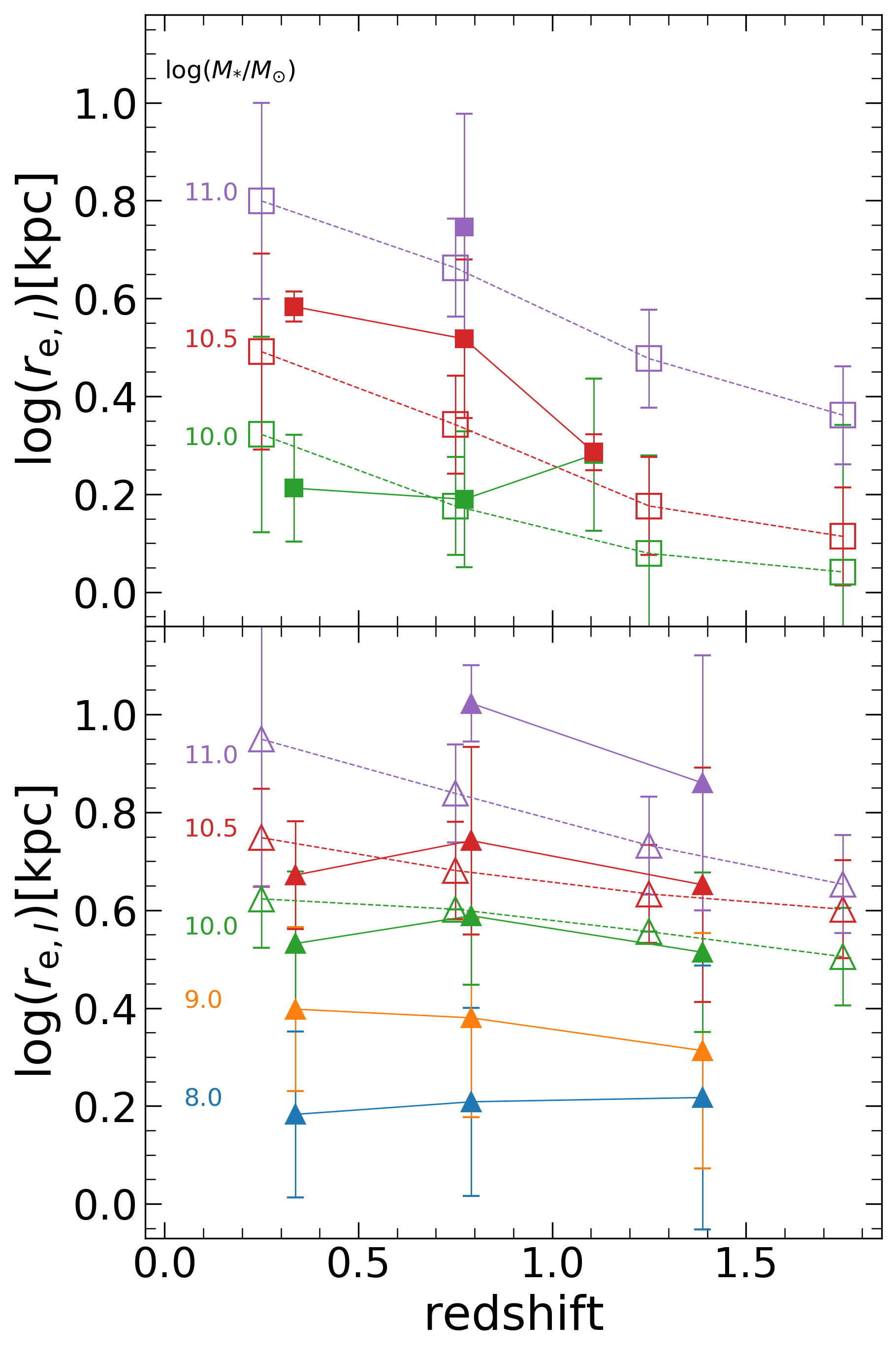}
        \caption[]{{Median size evolution with redshift, the $r_{\rm e}$--$z$ relation. Median sizes are shown for fixed stellar mass bins centered at \logmass$\,=\,8, 9, 10, 10.5, {\rm and}\ 11$ with colours indicated in the Figure (bin width of 0.5 dex.). We note that only three of the most massive bins are shown for ET for each volume (top panel). Filled symbols represent data from this work, while open markers shows data from \citet{Mowla2019ApJ...880...57M}, their Table 3 (including errors). Our error bars represent the median absolute deviation of the data in each bin. We plot median sizes for bins that have more than three sources.}}
        \label{fig:mass_evolution}
\end{figure}

\begin{figure}[t!]
        \centering
        \includegraphics[width=\linewidth]{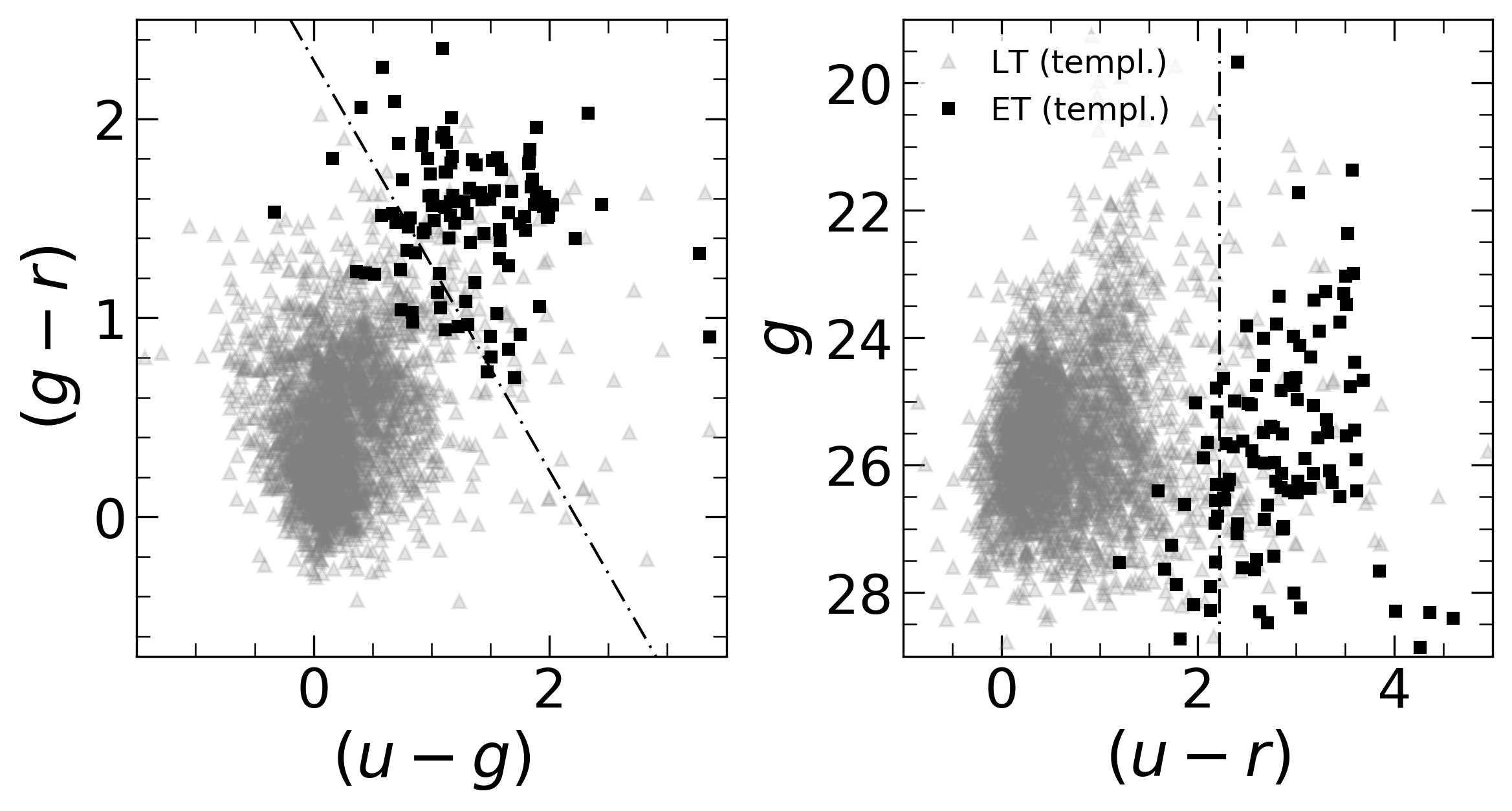}
        \caption[]{ET-LT colour separation from \cite{Strateva2001AJ....122.1861S} using data from this work. Gray triangles and black squares represents LT and ET galaxies classified via templates. Dot-dashed line represent \colourUR\,=\,2.22.}
        \label{fig:colour_strateva}
\end{figure}

\cite{Strateva2001AJ....122.1861S} demonstrated that a \colourUR\,=\,2.22 colour-cut separation can be useful for a broad segregation of ET from LT. They used a sample with more than 140\,000 galaxies from the SDSS survey down to magnitude $g=21$ and $z\,<\,0.4$. Even though this particular method does not directly use the parameters derived in this work, it is interesting to test its performance using the OTELO catalogue. This method uses parameters that are relatively easy to obtain, namely, the observed magnitudes in the \textit{g}, \textit{r,} and \textit{u} bands, which are available also in the OTELO catalogue. Figure \ref{fig:colour_strateva} shows how this method is able to separate sources in the \morphsample\ up to \photz$\,=\,2$. Nearly 81\% (97\%) of ET (LT) from the \morphsample\ are correctly separated, namely, with \colourUR$\,\geq\,$2.22 (\textless\,2.22). The reliability (defined as in \citealt{Strateva2001AJ....122.1861S}) of ET (LT) selection for this method is of 54\% (99\%). As compared to values from \cite{Strateva2001AJ....122.1861S}, their ET (LT) completeness and reliability are 98\% (72\%) and 83\% (96\%) for the spectroscopic sample and  80\% (66\%) and 62\% (83\%) for visually classified galaxies, respectively (see their Table 2 and 3). Although the completeness of our ET selection using this method  decreases slightly, our sample extends to a much higher redshift and is deeper in magnitude than both samples presented in \cite{Strateva2001AJ....122.1861S}. Despite this, an appropriate \colourUR\ colour-cut separation could be used as a fair proxy to a ET-LT segregation in OTELO data. 

The stellar mass-size relations for ET and LT (Figure  \ref{fig:mass_size}) in our sample, regardless of the separation in redshift bins, are in agreement with the general trends at low redshift reported in the recent literature \citep{Shen2003MNRAS.343..978S,LangeGAMA_Mass_Re_2015}. According to these authors, and leaving aside the data scattering, the MSR for ET galaxies is steeper than the corresponding to LT, { as confirmed in this work. Furthermore,} the stellar mass distributions show a sort of bi-modality, with LT galaxies being less massive than ET sources (in our case, with median values of \logmass\,$\,=\,8.8$ and 10.12, respectively). 
Finally, according to the results given in Section \ref{sec:morphological_analysis} we cannot draw any conclusion about the possible evolution of the MSR in both types, {considering stellar-mass range studied (Fig. \ref{fig:mass_size}). Recent studies of the MSR in similar redshift range as studied in this work \citep[e.g.][and references therein]{vanderWel2014ApJ...788...28V,Roy2018,Mowla2019ApJ...880...57M} point out to a median  size evolution of galaxies with redshift ($r_{\rm e}$--$z$). In particular, \citet{Mowla2019ApJ...880...57M} used $I$-band \hstasc\ images to quantify the $r_{\rm e}$--$z$ relation for COSMOS-DASH survey, that is, the same instrument and filter of the \hst\ as in this work. The use of the same photometric band is especially important because of the claimed dependence of size on the wavelength  at which the measurements are done \citep[e.g.][]{Kelvin_2012MNRAS.421.1007K}. Since the MSR presented by \citet{Mowla2019ApJ...880...57M} do not match stellar-mass ranges studied here, we compare the $r_{\rm e}$--$z$ for three fixed mass of \logmass$\,= 10, 10.5, {\rm and}, 11 $ and, furthermore, including lower-mass bins of \logmass\,$= 8$ and $9$ for the LT galaxies from this work. In Figure \ref{fig:mass_evolution}, we show the results of this comparison. In this figure, we only plotted median size values for bins where we have more than three objects (we note that for the ET most massive stellar-mass bin, we plotted only the intermediate redshift bin). This ensures a more robust comparison with previous works.
Considering, the $r_{\rm e}$--$z$ relation for fixed stellar-mass of \logmass$\,=10.5$ (red lines and points in Fig. \ref{fig:mass_evolution}) for which OTELO have sufficient statistics in both morphological types, we can notice that our results are consistent with \citet{Mowla2019ApJ...880...57M}: ET galaxies present steeper median size evolution, as compared to LT population. Furthermore, we show the the median size evolution for LT galaxies in the previously unexplored mass regime, namely, for \logmass\ of 8, and 9. These are two mass bins where OTELO do have sufficient number of sources in all redshifts (between $\sim$ 50, and $\sim$ 450). We find very mild evolution of median size for these masses, which is compatible with the scenario of passive evolution for this type of galaxy \citep[e.g.][]{vanderWel2014ApJ...788...28V,Mowla2019ApJ...880...57M}. Generally speaking, and taking into account the biases introduced in our selection process (under-representation of high-mass end), our results are in agreement with previous findings on MSR, as well as on the median size evolution $r_{\rm e}$--$z$.}

\section{Morphological catalogue}
\label{sec:morphological_catalogue}

\begin{figure*}
\centering
\subfloat[id: 1003]{
  \includegraphics[width=\columnwidth]{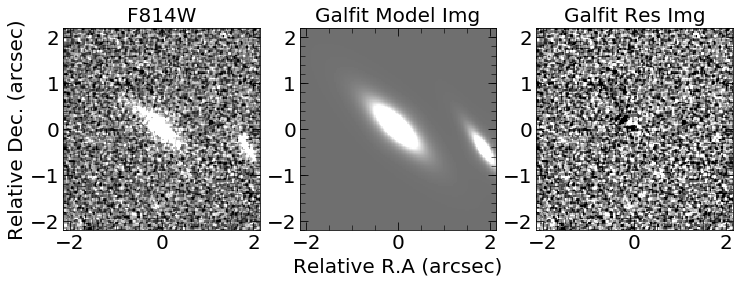}
}
\subfloat[id: 279]{
  \includegraphics[width=\columnwidth]{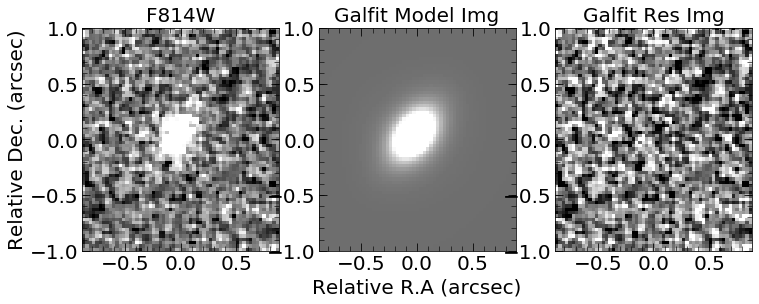}
}
\hspace{0mm}
\subfloat[id: 5691]{
  \includegraphics[width=\columnwidth]{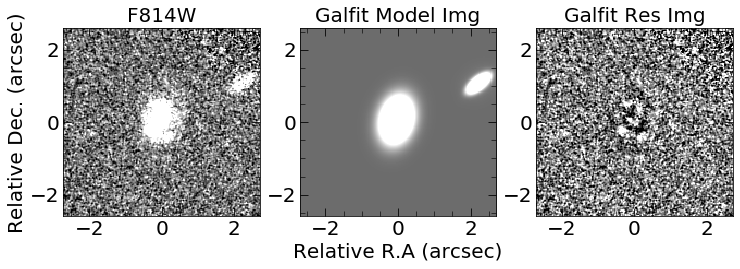}
}
\subfloat[id: 6176]{  
  \includegraphics[width=\columnwidth]{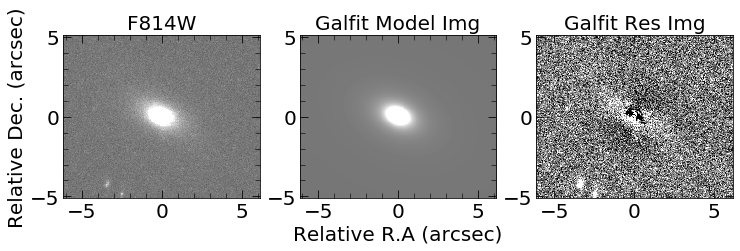}
}
\hspace{0mm}
\subfloat[id: 7249]{
  \includegraphics[width=\columnwidth]{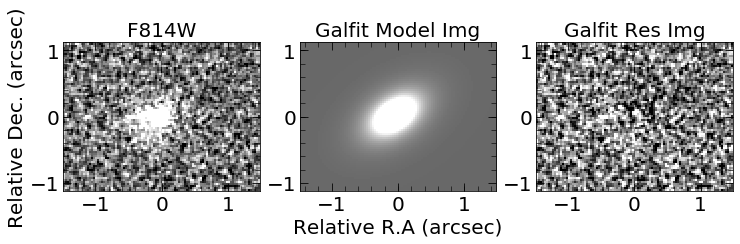}
}
\subfloat[id: 10555]{
  \includegraphics[width=\columnwidth]{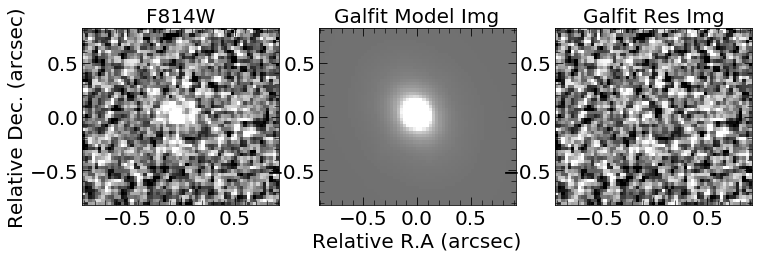}
}
\caption{Examples of LT (left column) and ET (right column) sources. For each source, we show (from left to right) the original \hstasc-I band image, \galfitm\ model and residual image. LT and ET classification is obtained from SED fitting to the templates, as described in Section \ref{sec:otelo_catalogue}}
\label{fig:sources}
\end{figure*}

Together with this work, we are making  the morphological catalogue public, along with the aforementioned parameters. {The catalogue can be retrieved from our website or from the CDS data-base\footnote{via anonymous ftp to cdsarc.u-strasbg.fr (130.79.128.5) or via \url{http://cdsweb.u-strasbg.fr/cgi-bin/qcat?J/A+A/}}}. Table \ref{tab:catalogue} provides the description of parameters in the catalogue. The unique the OTELO object number \texttt{idobj} can be used to match results from this work with the OTELO catalogue. The catalogue presented in this work consists of the G2 sample (as defined in Section \ref{sec:sample_selection}) with 8813 detections in \hstasc\ \textit{I}-band image, namely, all the sources with a successful fit from \GALA\ (not necessarily meaningful). The flag \texttt{flag\_good} ($=1$) indicates the meaningful selection (6780). The derived parameters from \GALA\ were measured in both available bands, namely, \textit{V} (F606W) and \textit{I} (F814W) from \hstasc. The \morphsample\ (studied in this work) can be easily recovered using $0\,<$ \texttt{z\_phot} $<\,2$. These are sources with 'good' \GALA\ output parameters and with individual match to the OTELO catalogue (the closest in the case of multiple matches). For these sources, we include several parameters from the OTELO catalogue (unique OTELO id \texttt{idobj}, selected photometric redshift \texttt{z\_phot}, and \texttt{template} associated with selected redshift; see Section \ref{sec:otelo_catalogue}) and the stellar mass estimation from \citet{Nadolny}. The multiple matches can be identified by \texttt{GroupID} and \texttt{GroupSize}, however only the closest to the OTELO catalogue position is indicated, for sake of clarity. In Figure \ref{fig:sources} we show images from the fitting process with original, model, and residual images for several sources.

\section{Conclusions}
\label{sec:conclusions}
In this paper, we aim to present the morphological catalogue of 8813 sources detected in \hstasc\ $I$-band in the framework of the OTELO survey. This catalogue contains the multi-wavelength morphological parameters of a single-S\'ersic model fitted to \hstasc\ $V$- and $I$-bands employing \GALA, as well as stellar masses. The unique combination of this morphological and the OTELO catalogues provides a valuable tool for the study of different aspects of galaxy evolution. 

Using OTELO's ET-LT galaxy classification via templates, we examined some of the methods found in the literature using the derived parameters. A rigorous sample selection assures the exact correspondence of data from the OTELO catalogue to morphological parameters obtained from the high resolution images. We found great similarities in the results with regard to previous works of ET and LT separation in terms of the S\'ersic index \sersic, ratio of S\'ersic index in $I$- and $V$-band \N, and observed colour \colourUR. Furthermore, we also tested an independent classification method which uses only the observed colours, described in \cite{Strateva2001AJ....122.1861S}. A general agreement was found despite its own reliability.

Due to the statistical similarities among the ET-LT separation using methods employed in the low-$z$\ Universe, we found no evidence of evolution for the studied parameters. This is also confirmed in the case of the MSR relation, which we found to closely follow the local MSR from \citet[][\z\,<\,0.1]{LangeGAMA_Mass_Re_2015}, as shown in Figure \ref{fig:mass_size}. We note, however, that our selection process does indeed bias the sample, and in the case of MSR, we lose a portion of the massive (\logmass\,>\,11) ET galaxies. {This bias towards less massive galaxies in our sample is evident when comparing our results to the  sample obtained
with MSR of more massive (\logmass\,>\,11.3) sources at the same redshift bins \citep{Mowla2019ApJ...880...57M}. In particular, the MSR for ET galaxies for the highest redshift bin is not sufficiently sampled, resulting in an offset of $\sim\,$0.4 dex in size. This is also reflected in the errors of the power--law parameters fitted for this particular sub-sample. We also investigated the median size evolution since $z=2$, finding a good agreement with the recent study of \citet{Mowla2019ApJ...880...57M}: the median size of ET galaxies evolves more steeply than the median size of LT for a given stellar mass.} In any case, these results corroborate the fact that our sample is composed of field galaxies. Thus, we are in the position to make comparisons between cluster versus field galaxies -- which is to make up the scope of a future work. In this context, we will make use of the GLACE survey (\citealt{glace2015A&A...578A..30S}), whose design and purpose are very closely related to those of the OTELO survey.

Several scientific cases are under study by our team using this dataset. These are: a comparison of sources with and without emission lines detected in OTELO; using machine learning techniques to improve ET-LT separation through colour proxies, morphology, and redshift (e.g. using the catalogue from this work with OTELO photometry and redshift in \citealt{JAD}); a comparison of cluster versus field galaxies; and, finally, a study of compact galaxies and ET galaxies with emission lines. The latter two are already in preparation. 

Together with this paper, we provide a public morphological catalogue with 8813 entries as described in Section \ref{sec:morphological_catalogue} and in Table \ref{tab:catalogue}. The sources studied in this article, along with additional information (stellar masses, \photz,\ and morphological classification via templates) can be easily recovered from the catalogue.

\begin{acknowledgements}
This paper is dedicated to the memory of our dear friend and colleague Hector Casta\~neda, unfortunately died on 19 Nov 2020.
This work was supported by the project Evolution of Galaxies, of reference AYA2014-58861-C3-1-P and AYA2017-88007-C3-1-P, within the "Programa estatal de fomento de la investigaci\'on cient\'ifica y t\'ecnica de excelencia del Plan Estatal de Investigaci\'on Cient\'ifica y T\'ecnica y de Innovaci\'on (2013-2016)" of the "Agencia Estatal de Investigaci\'on del Ministerio de Ciencia, Innovaci\'on y Universidades", and co-financed by the FEDER "Fondo Europeo de Desarrollo Regional". APG, RPM and MSP was supported by AYA2017-88007-C3-2-P. APG and MC are also supported by the Spanish State Research Agency grant MDM-2017-0737 (Unidad de Excelencia Mar\'ia de Maeztu  CAB). JAD is grateful for the support from the UNAM-DGAPA-PASPA 2019 program, and the kind hospitality of the IAC.

Based on observations made with the Gran Telescopio Canarias (GTC), installed in the Spanish Observatorio del Roque de los Muchachos of the Instituto de Astrof\'isica de Canarias, on the island of La Palma.

This paper made use of the IAC Supercomputing facility HTCondor (\url{http://research.cs.wisc.edu/htcondor/}).

This paper made use of the \cite{Wright2006PASP..118.1711W} cosmological calculator.

Based on observations obtained with MegaPrime/MegaCam, a joint project of CFHT and CEA/IRFU, at the Canada-France-Hawaii Telescope (CFHT) which is operated by the National Research Council (NRC) of Canada, the Institut National des Science de l'Univers of the Centre National de la Recherche Scientifique (CNRS) of France, and the University of Hawaii. This work is based in part on data products produced at Terapix available at the Canadian Astronomy Data Centre as part of the Canada-France-Hawaii Telescope Legacy Survey, a collaborative project of NRC and CNRS.
M.A.L.L is a DARK-Carlsberg Foundation Fellow (Semper Ardens project CF15-0384).

MP acknowledges financial financial support from the  Ethiopian  Space  Science  and  Technology  Institute  (ESSTI)  under  the  Ethiopian  Ministry  of Innovation  and  Technology  (MInT), and of  the  Spanish  MEC  undergrant  AYA2016-76682-C3-1-P and PID2019-106027GB-C41  and  financial  support  from the State Agency for Research of the Spanish MCIU through the "Center of Excellence Severo Ochoa" award to the Instituto de Astrof\'isica de Andaluc\'a (SEV-2017-0709).

\end{acknowledgements}

\begin{center}
\begin{table*}
      \caption[]{Column description of the morphological catalogue.}
         \label{tab:catalogue}
\begin{tabular}{ll}
\hline\hline \vspace{4pt}
Parameter & Description \\
\hline
\texttt{NUMBER}                                 & \se\ object number in \hstasc\ $I$-band \GALA\ catalogue \\
\texttt{ALPHA}, \texttt{DELTA}  & Equatorial coordinates (J2000) of the object in \hstasc\ $I$-band image \\
\texttt{MAG\_GF\_X}$^{(a,b)}$   & Output \texttt{X}-band magnitude\\
\texttt{MAGERR\_GF\_X}                  & Error on output \texttt{X}-band magnitude \\
\texttt{RE\_GF\_X}                              & Output \texttt{X}-band effective radius  \\
\texttt{REERR\_GF\_X}                   & Error on output \texttt{X}-band effective radius  \\
\texttt{N\_GF\_X}                               & Output \texttt{X}-band S\'ersic index \sersic\\
\texttt{NERR\_GF\_X}                    & Error on output \texttt{X}-band S\'ersic index \sersic \\
\texttt{Q\_GF\_X}                               & Output \texttt{X}-band axis-ratio \\
\texttt{QERR\_GF\_X}                    & Error on output \texttt{X}-band  axis-ratio \\
\texttt{PA\_GF\_X}                              & Output \texttt{X}-band position angle\\
\texttt{PAERR\_GF\_X}                   & Error on output \texttt{X}-band position angle\\
\texttt{FLUX\_RADIUS\_X\_Y}$^{(c)}$             & \se\ flux radius for \texttt{Y}\% of the total flux in \texttt{X}-band \\
\texttt{MAG\_AUTO\_X}                   & \se\ \texttt{X}-band magnitude \\
\texttt{MAGERR\_AUTO\_X}                & Error on \se\ \texttt{X}-band magnitude  \\
\texttt{flag\_good}             & Flag indicating "good" \GALA\ results ($=$1; see Section \ref{sec:sample_selection}) \\
\hline
\texttt{idobj}$^{(d)}$          & OTELO object number in raw catalogue \\
\texttt{GroupID}                        & Index of the group of sources matched to the same OTELO \texttt{idobj}  \\
\texttt{GroupSize}                      & Number of sources \texttt{NUMBER} in given \texttt{GroupID} \\
\texttt{Separation}             & Separation of \texttt{NUMBER} in arcseconds from OTELO source \texttt{idobj} \\
\texttt{z\_phot}                & Photometric redshift \photz\ selected from OTELO catalogue \\
\texttt{z\_phot\_err}   & Error on \photz\ selected from OTELO catalogue \\
\texttt{template}               & Template associated with \photz\ solution used in this work from OTELO catalogue \\
\texttt{LogM}                   & Stellar mass estimated using template associated with \photz  \\
\texttt{LogM\_err}                              & Error on stellar mass \\
\hline
\end{tabular}
\tablefoot{\\ $^{(a)}$ where \texttt{GF} is \galfitm\ output value. \\
$^{(b)}$ where \texttt{X} is V or I (\hstasc\ F606W or F814W, respectively). \\
$^{(c)}$ where \texttt{Y} is 20, 30, 50, 80 or 90 (\% of the total flux in given radius). \\
$^{(d)}$ \texttt{idobj} - \texttt{LogM\_err} only sources with individual match with OTELO or the closest of the multiple matches ({multiple main}); total of 3658.}
\end{table*}
\end{center}

%
%
\bibliographystyle{aa} 
\bibliography{bib} 

\begin{thebibliography}{76}
\expandafter\ifx\csname natexlab\endcsname\relax\def\natexlab#1{#1}\fi

\bibitem[{{Abraham} {et~al.}(1994){Abraham}, {Valdes}, {Yee}, \& {van den
  Bergh}}]{Abraham1994ApJ...432...75A}
{Abraham}, R.~G., {Valdes}, F., {Yee}, H.~K.~C., \& {van den Bergh}, S. 1994,
  \apj, 432, 75

\bibitem[{{Abraham} {et~al.}(1996){Abraham}, {van den Bergh}, {Glazebrook},
  {Ellis}, {Santiago}, {Surma}, \& {Griffiths}}]{Abraham1996ApJS..107....1A}
{Abraham}, R.~G., {van den Bergh}, S., {Glazebrook}, K., {et~al.} 1996, \apjs,
  107, 1

\bibitem[{{Abraham} {et~al.}(2003){Abraham}, {van den Bergh}, \&
  {Nair}}]{Abraham2003ApJ...588..218A}
{Abraham}, R.~G., {van den Bergh}, S., \& {Nair}, P. 2003, \apj, 588, 218

\bibitem[{{Andrae} {et~al.}(2011b){Andrae}, {Jahnke}, \&
  {Melchior}}]{Andrae2011MNRAS.411..385A}
{Andrae}, R., {Jahnke}, K., \& {Melchior}, P. 2011b, \mnras, 411, 385

\bibitem[{{Andrae} {et~al.}(2011a){Andrae}, {Melchior}, \&
  {Jahnke}}]{Andrae2011MNRAS.417.2465A}
{Andrae}, R., {Melchior}, P., \& {Jahnke}, K. 2011a, \mnras, 417, 2465

\bibitem[{{Arnouts} {et~al.}(1999){Arnouts}, {Cristiani}, {Moscardini},
  {Matarrese}, {Lucchin}, {Fontana}, \& {Giallongo}}]{Arnouts1999}
{Arnouts}, S., {Cristiani}, S., {Moscardini}, L., {et~al.} 1999, \mnras, 310,
  540

\bibitem[{{Baldry} {et~al.}(2004){Baldry}, {Glazebrook}, {Brinkmann},
  {Ivezi{\'c}}, {Lupton}, {Nichol}, \& {Szalay}}]{Baldry2004ApJ...600..681B}
{Baldry}, I.~K., {Glazebrook}, K., {Brinkmann}, J., {et~al.} 2004, \apj, 600,
  681

\bibitem[{{Barden} {et~al.}(2012){Barden}, {H{\"a}u{\ss}ler}, {Peng},
  {McIntosh}, \& {Guo}}]{Barden2012MNRAS.422..449B}
{Barden}, M., {H{\"a}u{\ss}ler}, B., {Peng}, C.~Y., {McIntosh}, D.~H., \&
  {Guo}, Y. 2012, \mnras, 422, 449

\bibitem[{{Barden} {et~al.}(2005){Barden}, {Rix}, {Somerville}, {Bell},
  {H{\"a}u{\ss}ler}, {Peng}, {Borch}, {Beckwith}, {Caldwell}, {Heymans},
  {Jahnke}, {Jogee}, {McIntosh}, {Meisenheimer}, {S{\'a}nchez}, {Wisotzki}, \&
  {Wolf}}]{Barden2005ApJ...635..959B}
{Barden}, M., {Rix}, H.-W., {Somerville}, R.~S., {et~al.} 2005, \apj, 635, 959

\bibitem[{{Bershady} {et~al.}(2000){Bershady}, {Jangren}, \&
  {Conselice}}]{Bershady2000AJ....119.2645B}
{Bershady}, M.~A., {Jangren}, A., \& {Conselice}, C.~J. 2000, \aj, 119, 2645

\bibitem[{{Bertin} \& {Arnouts}(1996)}]{SEx1996}
{Bertin}, E. \& {Arnouts}, S. 1996, \aaps, 117, 393

\bibitem[{{Bertin} {et~al.}(2002){Bertin}, {Mellier}, {Radovich}, {Missonnier},
  {Didelon}, \& {Morin}}]{Bertin2002}
{Bertin}, E., {Mellier}, Y., {Radovich}, M., {et~al.} 2002, in Astronomical
  Society of the Pacific Conference Series, Vol. 281, Astronomical Data
  Analysis Software and Systems XI, ed. D.~A. {Bohlender}, D.~{Durand}, \&
  T.~H. {Handley}, 228

\bibitem[{{Bongiovanni} {et~al.}(2019){Bongiovanni}, {Ram{\'o}n-P{\'e}rez},
  {P{\'e}rez Garc{\'\i}a}, {Cepa}, {Cervi{\~n}o}, {Nadolny}, {P{\'e}rez
  Mart{\'\i}nez}, {Alfaro}, {Casta{\~n}eda}, {de Diego}, {Ederoclite},
  {Fern{\'a}ndez-Lorenzo}, {Gallego}, {Gonz{\'a}lez}, {Gonz{\'a}lez-Serrano},
  {Lara-L{\'o}pez}, {Oteo G{\'o}mez}, {Padilla Torres}, {Pintos-Castro},
  {Povi{\'c}}, {S{\'a}nchez-Portal}, {Jones}, {Bland-Hawthorn}, \&
  {Cabrera-Lavers}}]{OteloI}
{Bongiovanni}, {\'A}., {Ram{\'o}n-P{\'e}rez}, M., {P{\'e}rez Garc{\'\i}a},
  A.~M., {et~al.} 2019, \aap, 631, A9

\bibitem[{{Brammer} {et~al.}(2012){Brammer}, {van Dokkum}, {Franx},
  {Fumagalli}, {Patel}, {Rix}, {Skelton}, {Kriek}, {Nelson}, {Schmidt},
  {Bezanson}, {da Cunha}, {Erb}, {Fan}, {F{\"o}rster Schreiber}, {Illingworth},
  {Labb{\'e}}, {Leja}, {Lundgren}, {Magee}, {Marchesini}, {McCarthy},
  {Momcheva}, {Muzzin}, {Quadri}, {Steidel}, {Tal}, {Wake}, {Whitaker}, \&
  {Williams}}]{CANDELS_AEGIS2012ApJS..200...13B}
{Brammer}, G.~B., {van Dokkum}, P.~G., {Franx}, M., {et~al.} 2012, \apjs, 200,
  13

\bibitem[{{Cepa} {et~al.}(2003){Cepa}, {Aguiar-Gonzalez}, {Bland -Hawthorn},
  {Castaneda}, {Cobos}, {Correa}, {Espejo}, {Fragoso-Lopez}, {Fuentes},
  {Gigante}, {Gonzalez}, {Gonzalez-Escalera}, {Gonzalez-Serrano},
  {Joven-Alvarez}, {Lopez-Ruiz}, {Militello}, {Cano}, {Perez}, {Perez},
  {Rasilla}, {Sanchez}, \& {Tejada}}]{2003SPIE.4841.1739C}
{Cepa}, J., {Aguiar-Gonzalez}, M., {Bland -Hawthorn}, J., {et~al.} 2003, in
  Society of Photo-Optical Instrumentation Engineers (SPIE) Conference Series,
  Vol. 4841, \procspie, ed. M.~{Iye} \& A.~F.~M. {Moorwood} (SPIE), 1739--1749

\bibitem[{{Ciotti}(1991)}]{Ciotti1991A&A...249...99C}
{Ciotti}, L. 1991, \aap, 249, 99

\bibitem[{{Coleman} {et~al.}(1980){Coleman}, {Wu}, \& {Weedman}}]{Coleman1980}
{Coleman}, G.~D., {Wu}, C.-C., \& {Weedman}, D.~W. 1980, \apjs, 43, 393

\bibitem[{{Conselice} {et~al.}(2000){Conselice}, {Bershady}, \&
  {Jangren}}]{Conselice2000ApJ...529..886C}
{Conselice}, C.~J., {Bershady}, M.~A., \& {Jangren}, A. 2000, \apj, 529, 886

\bibitem[{{de Diego} {et~al.}(2020){de Diego}, {Nadolny}, {Bongiovanni},
  {Cepa}, {Povi{\'c}}, {P{\'e}rez Garc{\'\i}a}, {Padilla Torres},
  {Lara-L{\'o}pez}, {Cervi{\~n}o}, {P{\'e}rez Mart{\'\i}nez}, {Alfaro},
  {Casta{\~n}eda}, {Fern{\'a}ndez-Lorenzo}, {Gallego}, {Gonz{\'a}lez},
  {Gonz{\'a}lez-Serrano}, {Pintos-Castro}, {S{\'a}nchez-Portal}, {Cedr{\'e}s},
  {Gonz{\'a}lez-Otero}, {Heath Jones}, \& {Bland-Hawthorn}}]{JAD}
{de Diego}, J.~A., {Nadolny}, J., {Bongiovanni}, {\'A}., {et~al.} 2020, \aap,
  638, A134

\bibitem[{{de Souza} {et~al.}(2004){de Souza}, {Gadotti}, \& {dos
  Anjos}}]{deSouza2004ApJS..153..411D}
{de Souza}, R.~E., {Gadotti}, D.~A., \& {dos Anjos}, S. 2004, \apjs, 153, 411

\bibitem[{{de Vaucouleurs}(1959)}]{Vaucouleurs1959HDP....53..275D}
{de Vaucouleurs}, G. 1959, Handbuch der Physik, 53, 275

\bibitem[{{Dressler} {et~al.}(1987){Dressler}, {Faber}, {Burstein}, {Davies},
  {Lynden-Bell}, {Terlevich}, \& {Wegner}}]{Dressler1987ApJ...313L..37D}
{Dressler}, A., {Faber}, S.~M., {Burstein}, D., {et~al.} 1987, \apjl, 313, L37

\bibitem[{{Freeman}(1970)}]{Freeman1970ApJ...160..811F}
{Freeman}, K.~C. 1970, \apj, 160, 811

\bibitem[{{Graham} {et~al.}(2005){Graham}, {Driver}, {Petrosian}, {Conselice},
  {Bershady}, {Crawford}, \& {Goto}}]{Graham2005AJ....130.1535G}
{Graham}, A.~W., {Driver}, S.~P., {Petrosian}, V., {et~al.} 2005, \aj, 130,
  1535

\bibitem[{{Griffith} {et~al.}(2012){Griffith}, {Cooper}, {Newman}, {Moustakas},
  {Stern}, {Comerford}, {Davis}, {Lotz}, {Barden}, {Conselice}, {Capak},
  {Faber}, {Kirkpatrick}, {Koekemoer}, {Koo}, {Noeske}, {Scoville}, {Sheth},
  {Shopbell}, {Willmer}, \& {Weiner}}]{Griffith2012ApJS..200....9G}
{Griffith}, R.~L., {Cooper}, M.~C., {Newman}, J.~A., {et~al.} 2012, \apjs, 200,
  9

\bibitem[{{H{\"a}u{\ss}ler} {et~al.}(2013){H{\"a}u{\ss}ler}, {Bamford}, {Vika},
  {Rojas}, {Barden}, {Kelvin}, {Alpaslan}, {Robotham}, {Driver}, {Baldry},
  {Brough}, {Hopkins}, {Liske}, {Nichol}, {Popescu}, \&
  {Tuffs}}]{galapagos2013MNRAS.430..330H}
{H{\"a}u{\ss}ler}, B., {Bamford}, S.~P., {Vika}, M., {et~al.} 2013, \mnras,
  430, 330

\bibitem[{{H{\"a}ussler} {et~al.}(2007){H{\"a}ussler}, {McIntosh}, {Barden},
  {Bell}, {Rix}, {Borch}, {Beckwith}, {Caldwell}, {Heymans}, {Jahnke}, {Jogee},
  {Koposov}, {Meisenheimer}, {S{\'a}nchez}, {Somerville}, {Wisotzki}, \&
  {Wolf}}]{Haussler2007}
{H{\"a}ussler}, B., {McIntosh}, D.~H., {Barden}, M., {et~al.} 2007, \apjs, 172,
  615

\bibitem[{{Hubble}(1925)}]{Hubble1925ApJ....62..409H}
{Hubble}, E.~P. 1925, \apj, 62, 409

\bibitem[{{Hubble}(1936)}]{Hubble1936rene.book.....H}
{Hubble}, E.~P. 1936, {Realm of the Nebulae} (Yale University Press)

\bibitem[{{Huertas-Company} {et~al.}(2015){Huertas-Company}, {Gravet},
  {Cabrera-Vives}, {P{\'e}rez-Gonz{\'a}lez}, {Kartaltepe}, {Barro}, {Bernardi},
  {Mei}, {Shankar}, {Dimauro}, {Bell}, {Kocevski}, {Koo}, {Faber}, \&
  {Mcintosh}}]{Huertas2015ApJS..221....8H}
{Huertas-Company}, M., {Gravet}, R., {Cabrera-Vives}, G., {et~al.} 2015, \apjs,
  221, 8

\bibitem[{{Huertas-Company} {et~al.}(2008){Huertas-Company}, {Rouan}, {Tasca},
  {Soucail}, \& {Le F{\`e}vre}}]{Huertas2008A&A...478..971H}
{Huertas-Company}, M., {Rouan}, D., {Tasca}, L., {Soucail}, G., \& {Le
  F{\`e}vre}, O. 2008, \aap, 478, 971

\bibitem[{{Ilbert} {et~al.}(2006){Ilbert}, {Arnouts}, {McCracken},
  {Bolzonella}, {Bertin}, {Le F{\`e}vre}, {Mellier}, {Zamorani}, {Pell{\`o}},
  {Iovino}, {Tresse}, {Le Brun}, {Bottini}, {Garilli}, {Maccagni}, {Picat},
  {Scaramella}, {Scodeggio}, {Vettolani}, {Zanichelli}, {Adami}, {Bardelli},
  {Cappi}, {Charlot}, {Ciliegi}, {Contini}, {Cucciati}, {Foucaud}, {Franzetti},
  {Gavignaud}, {Guzzo}, {Marano}, {Marinoni}, {Mazure}, {Meneux}, {Merighi},
  {Paltani}, {Pollo}, {Pozzetti}, {Radovich}, {Zucca}, {Bondi}, {Bongiorno},
  {Busarello}, {de La Torre}, {Gregorini}, {Lamareille}, {Mathez}, {Merluzzi},
  {Ripepi}, {Rizzo}, \& {Vergani}}]{Ilbert2006}
{Ilbert}, O., {Arnouts}, S., {McCracken}, H.~J., {et~al.} 2006, \aap, 457, 841

\bibitem[{{Jim{\'e}nez-Teja} \&
  {Ben{\'\i}tez}(2012)}]{JimenezTeja2012ApJ...745..150J}
{Jim{\'e}nez-Teja}, Y. \& {Ben{\'\i}tez}, N. 2012, \apj, 745, 150

\bibitem[{{Kauffmann} {et~al.}(2003){Kauffmann}, {Heckman}, {White}, {Charlot},
  {Tremonti}, {Peng}, {Seibert}, {Brinkmann}, {Nichol}, {SubbaRao}, \&
  {York}}]{Kauffmann2003MNRAS.341...54K}
{Kauffmann}, G., {Heckman}, T.~M., {White}, S. D.~M., {et~al.} 2003, \mnras,
  341, 54

\bibitem[{Kelly \& McKay(2005)}]{Kelly_2005}
Kelly, B.~C. \& McKay, T.~A. 2005, The Astronomical Journal, 129, 1287

\bibitem[{{Kelvin} {et~al.}(2012){Kelvin}, {Driver}, {Robotham}, {Hill},
  {Alpaslan}, {Baldry}, {Bamford}, {Bland-Hawthorn}, {Brough}, {Graham},
  {H{\"a}ussler}, {Hopkins}, {Liske}, {Loveday}, {Norberg}, {Phillipps},
  {Popescu}, {Prescott}, {Taylor}, \& {Tuffs}}]{Kelvin_2012MNRAS.421.1007K}
{Kelvin}, L.~S., {Driver}, S.~P., {Robotham}, A. S.~G., {et~al.} 2012, \mnras,
  421, 1007

\bibitem[{{Kinney} {et~al.}(1996){Kinney}, {Calzetti}, {Bohlin}, {McQuade},
  {Storchi-Bergmann}, \& {Schmitt}}]{Kinney1996}
{Kinney}, A.~L., {Calzetti}, D., {Bohlin}, R.~C., {et~al.} 1996, \apj, 467, 38

\bibitem[{{Koekemoer} {et~al.}(2003){Koekemoer}, {Fruchter}, {Hook}, \&
  {Hack}}]{Koekemoer2003}
{Koekemoer}, A.~M., {Fruchter}, A.~S., {Hook}, R.~N., \& {Hack}, W. 2003, in
  HST Calibration Workshop : Hubble after the Installation of the ACS and the
  NICMOS Cooling System, ed. S.~{Arribas}, A.~{Koekemoer}, \& B.~{Whitmore},
  337

\bibitem[{{Kormendy}(1977)}]{Kormendy1977ApJ...218..333K}
{Kormendy}, J. 1977, \apj, 218, 333

\bibitem[{{Kormendy} \& {Bender}(1996)}]{Kormendy1996ApJ...464L.119K}
{Kormendy}, J. \& {Bender}, R. 1996, \apjl, 464, L119

\bibitem[{{Krywult} {et~al.}(2017){Krywult}, {Tasca}, {Pollo}, {Vergani},
  {Bolzonella}, {Davidzon}, {Iovino}, {Gargiulo}, {Haines}, {Scodeggio},
  {Guzzo}, {Zamorani}, {Garilli}, {Granett}, {de la Torre}, {Abbas}, {Adami},
  {Bottini}, {Cappi}, {Cucciati}, {Franzetti}, {Fritz}, {Le Brun}, {Le
  F{\`e}vre}, {Maccagni}, {Ma{\l}ek}, {Marulli}, {Polletta}, {Tojeiro},
  {Zanichelli}, {Arnouts}, {Bel}, {Branchini}, {Coupon}, {De Lucia}, {Ilbert},
  {McCracken}, {Moscardini}, \& {Takeuchi}}]{Krywult2017A&A...598A.120K}
{Krywult}, J., {Tasca}, L.~A.~M., {Pollo}, A., {et~al.} 2017, \aap, 598, A120

\bibitem[{{Kuchner} {et~al.}(2017){Kuchner}, {Ziegler}, {Verdugo}, {Bamford},
  \& {H{\"a}u{\ss}ler}}]{Kuchner2017A&A...604A..54K}
{Kuchner}, U., {Ziegler}, B., {Verdugo}, M., {Bamford}, S., \&
  {H{\"a}u{\ss}ler}, B. 2017, \aap, 604, A54

\bibitem[{{Lange} {et~al.}(2015){Lange}, {Driver}, {Robotham}, {Kelvin},
  {Graham}, {Alpaslan}, {Andrews}, {Baldry}, {Bamford}, {Bland-Hawthorn},
  {Brough}, {Cluver}, {Conselice}, {Davies}, {Haeussler}, {Konstantopoulos},
  {Loveday}, {Moffett}, {Norberg}, {Phillipps}, {Taylor},
  {L{\'o}pez-S{\'a}nchez}, \& {Wilkins}}]{LangeGAMA_Mass_Re_2015}
{Lange}, R., {Driver}, S.~P., {Robotham}, A. S.~G., {et~al.} 2015, \mnras, 447,
  2603

\bibitem[{{Le F{\`e}vre} {et~al.}(2005){Le F{\`e}vre}, {Vettolani}, {Garilli},
  {Tresse}, {Bottini}, {Le Brun}, {Maccagni}, {Picat}, {Scaramella},
  {Scodeggio}, {Zanichelli}, {Adami}, {Arnaboldi}, {Arnouts}, {Bardelli},
  {Bolzonella}, {Cappi}, {Charlot}, {Ciliegi}, {Contini}, {Foucaud},
  {Franzetti}, {Gavignaud}, {Guzzo}, {Ilbert}, {Iovino}, {McCracken}, {Marano},
  {Marinoni}, {Mathez}, {Mazure}, {Meneux}, {Merighi}, {Paltani}, {Pell{\`o}},
  {Pollo}, {Pozzetti}, {Radovich}, {Zamorani}, {Zucca}, {Bondi}, {Bongiorno},
  {Busarello}, {Lamareille}, {Mellier}, {Merluzzi}, {Ripepi}, \&
  {Rizzo}}]{VVDS2005A&A...439..845L}
{Le F{\`e}vre}, O., {Vettolani}, G., {Garilli}, B., {et~al.} 2005, \aap, 439,
  845

\bibitem[{{L{\'o}pez-Sanjuan} {et~al.}(2018){L{\'o}pez-Sanjuan},
  {D{\'{\i}}az-Garc{\'{\i}}a}, {Cenarro}, {Fern{\'a}ndez-Soto}, {Viironen},
  {Molino}, {Ben{\'{\i}}tez}, {Crist{\'o}bal-Hornillos}, {Moles}, {Varela},
  {Arnalte-Mur}, {Ascaso}, {Castander}, {Cervi{\~n}o}, {Gonz{\'a}lez Delgado},
  {Husillos}, {M{\'a}rquez}, {Masegosa}, {Del Olmo}, {Povi{\'c}}, \&
  {Perea}}]{LopezSanjuan2018}
{L{\'o}pez-Sanjuan}, C., {D{\'{\i}}az-Garc{\'{\i}}a}, L.~A., {Cenarro}, A.~J.,
  {et~al.} 2018, ArXiv e-prints [\eprint[arXiv]{1805.03609}]

\bibitem[{{L{\'o}pez-Sanjuan} {et~al.}(2012){L{\'o}pez-Sanjuan}, {Le
  F{\`e}vre}, {Ilbert}, {Tasca}, {Bridge}, {Cucciati}, {Kampczyk}, {Pozzetti},
  {Xu}, {Carollo}, {Contini}, {Kneib}, {Lilly}, {Mainieri}, {Renzini},
  {Sanders}, {Scodeggio}, {Scoville}, {Taniguchi}, {Zamorani}, {Aussel},
  {Bardelli}, {Bolzonella}, {Bongiorno}, {Capak}, {Caputi}, {de la Torre}, {de
  Ravel}, {Franzetti}, {Garilli}, {Iovino}, {Knobel}, {Kova{\v{c}}},
  {Lamareille}, {Le Borgne}, {Le Brun}, {Le Floc'h}, {Maier}, {McCracken},
  {Mignoli}, {Pell{\'o}}, {Peng}, {P{\'e}rez-Montero}, {Presotto},
  {Ricciardelli}, {Salvato}, {Silverman}, {Tanaka}, {Tresse}, {Vergani},
  {Zucca}, {Barnes}, {Bordoloi}, {Cappi}, {Cimatti}, {Coppa}, {Koekemoer},
  {Liu}, {Moresco}, {Nair}, {Oesch}, {Schawinski}, \&
  {Welikala}}]{LsJ_COSMOS_ET_2012A&A...548A...7L}
{L{\'o}pez-Sanjuan}, C., {Le F{\`e}vre}, O., {Ilbert}, O., {et~al.} 2012, \aap,
  548, A7

\bibitem[{{Lotz} {et~al.}(2004){Lotz}, {Primack}, \&
  {Madau}}]{Lotz2004AJ....128..163L}
{Lotz}, J.~M., {Primack}, J., \& {Madau}, P. 2004, \aj, 128, 163

\bibitem[{{Mahoro} {et~al.}(2019){Mahoro}, {Povi{\'c}}, {Nkundabakura},
  {Nyiransengiyumva}, \& {V{\"a}is{\"a}nen}}]{Mahoro2019MNRAS.485..452M}
{Mahoro}, A., {Povi{\'c}}, M., {Nkundabakura}, P., {Nyiransengiyumva}, B., \&
  {V{\"a}is{\"a}nen}, P. 2019, \mnras, 485, 452

\bibitem[{{Mowla} {et~al.}(2019){Mowla}, {van Dokkum}, {Brammer}, {Momcheva},
  {van der Wel}, {Whitaker}, {Nelson}, {Bezanson}, {Muzzin}, {Franx},
  {MacKenty}, {Leja}, {Kriek}, \& {Marchesini}}]{Mowla2019ApJ...880...57M}
{Mowla}, L.~A., {van Dokkum}, P., {Brammer}, G.~B., {et~al.} 2019, \apj, 880,
  57

\bibitem[{{Nadolny} {et~al.}(2020){Nadolny}, {Lara-L{\'o}pez}, {Cervi{\~n}o},
  {Bongiovanni}, {Cepa}, {de Diego}, {P{\'e}rez Garc{\'\i}a}, {P{\'e}rez
  Mart{\'\i}nez}, {S{\'a}nchez-Portal}, {Alfaro}, {Casta{\~n}eda}, {Gallego},
  {Gonz{\'a}lez}, {Gonz{\'a}lez-Serrano}, {Padilla Torres}, {Pintos-Castro}, \&
  {Povi{\'c}}}]{Nadolny}
{Nadolny}, J., {Lara-L{\'o}pez}, M.~A., {Cervi{\~n}o}, M., {et~al.} 2020, \aap,
  636, A84

\bibitem[{{Ngan} {et~al.}(2009){Ngan}, {van Waerbeke}, {Mahdavi}, {Heymans}, \&
  {Hoekstra}}]{Ngan2009MNRAS.396.1211N}
{Ngan}, W., {van Waerbeke}, L., {Mahdavi}, A., {Heymans}, C., \& {Hoekstra}, H.
  2009, \mnras, 396, 1211

\bibitem[{{Peng} {et~al.}(2002){Peng}, {Ho}, {Impey}, \& {Rix}}]{Galfit2002}
{Peng}, C.~Y., {Ho}, L.~C., {Impey}, C.~D., \& {Rix}, H.-W. 2002, \aj, 124, 266

\bibitem[{{Peng} {et~al.}(2010){Peng}, {Ho}, {Impey}, \&
  {Rix}}]{GALFIT_2010AJ....139.2097P}
{Peng}, C.~Y., {Ho}, L.~C., {Impey}, C.~D., \& {Rix}, H.-W. 2010, \aj, 139,
  2097

\bibitem[{{Planck Collaboration} {et~al.}(2016){Planck Collaboration}, {Ade},
  {Aghanim}, {Arnaud}, {Ashdown}, {Aumont}, {Baccigalupi}, {Banday},
  {Barreiro}, {Bartlett}, \& et~al.}]{Planck2016A&A...594A..13P}
{Planck Collaboration}, {Ade}, P.~A.~R., {Aghanim}, N., {et~al.} 2016, \aap,
  594, A13

\bibitem[{{Povi{\'c}} {et~al.}(2013){Povi{\'c}}, {Huertas-Company}, {Aguerri},
  {M{\'a}rquez}, {Masegosa}, {Husillos}, {Molino}, {Crist{\'o}bal-Hornillos},
  {Perea}, {Ben{\'\i}tez}, {Olmo}, {Fern{\'a}ndez-Soto}, {Jim{\'e}nez-Teja},
  {Moles}, {Alfaro}, {Aparicio-Villegas}, {Ascaso}, {Broadhurst},
  {Cabrera-Ca{\~n}o}, {Castander}, {Cepa}, {Fernand ez Lorenzo}, {Cervi{\~n}o},
  {Delgado}, {Infante}, {L{\'o}pez-Sanjuan}, {Mart{\'\i}nez}, {Matute}, {Oteo},
  {P{\'e}rez-Garc{\'\i}a}, {Prada}, \& {Quintana}}]{Povic2013MNRAS.435.3444P}
{Povi{\'c}}, M., {Huertas-Company}, M., {Aguerri}, J.~A.~L., {et~al.} 2013,
  \mnras, 435, 3444

\bibitem[{{Povi{\'c}} {et~al.}(2015){Povi{\'c}}, {M{\'a}rquez}, {Masegosa},
  {Perea}, {Olmo}, {Simpson}, {Aguerri}, {Ascaso}, {Jim{\'e}nez-Teja},
  {L{\'o}pez-Sanjuan}, {Molino}, {P{\'e}rez-Garc{\'\i}a}, {Viironen},
  {Husillos}, {Crist{\'o}bal-Hornillos}, {Caldwell}, {Ben{\'\i}tez}, {Alfaro},
  {Aparicio-Villegas}, {Broadhurst}, {Cabrera-Ca{\~n}o}, {Castander}, {Cepa},
  {Cervi{\~n}o}, {Fern{\'a}ndez-Soto}, {Delgado}, {Infante}, {Mart{\'\i}nez},
  {Moles}, {Prada}, \& {Quintana}}]{Povic2015MNRAS.453.1644P}
{Povi{\'c}}, M., {M{\'a}rquez}, I., {Masegosa}, J., {et~al.} 2015, \mnras, 453,
  1644

\bibitem[{{Povi{\'c}} {et~al.}(2009){Povi{\'c}}, {S{\'a}nchez-Portal},
  {P{\'e}rez Garc{\'\i}a}, {Bongiovanni}, {Cepa}, {Alfaro}, {Casta{\~n}eda},
  {Fern{\'a}ndez Lorenzo}, {Gallego}, {Gonz{\'a}lez-Serrano}, {Gonz{\'a}lez},
  \& {Lara-L{\'o}pez}}]{Povic2009ApJ...706..810P}
{Povi{\'c}}, M., {S{\'a}nchez-Portal}, M., {P{\'e}rez Garc{\'\i}a}, A.~M.,
  {et~al.} 2009, \apj, 706, 810

\bibitem[{{Povi{\'c}} {et~al.}(2012){Povi{\'c}}, {S{\'a}nchez-Portal},
  {P{\'e}rez Garc{\'\i}a}, {Bongiovanni}, {Cepa}, {Huertas-Company},
  {Lara-L{\'o}pez}, {Fern{\'a}ndez Lorenzo}, {Ederoclite}, {Alfaro},
  {Casta{\~n}eda}, {Gallego}, {Gonz{\'a}lez-Serrano}, \&
  {Gonz{\'a}lez}}]{Povic2012A&A...541A.118P}
{Povi{\'c}}, M., {S{\'a}nchez-Portal}, M., {P{\'e}rez Garc{\'\i}a}, A.~M.,
  {et~al.} 2012, \aap, 541, A118

\bibitem[{{Rix} {et~al.}(2004){Rix}, {Barden}, {Beckwith}, {Bell}, {Borch},
  {Caldwell}, {H{\"a}ussler}, {Jahnke}, {Jogee}, {McIntosh}, {Meisenheimer},
  {Peng}, {Sanchez}, {Somerville}, {Wisotzki}, \&
  {Wolf}}]{Rix2004ApJS..152..163R}
{Rix}, H.-W., {Barden}, M., {Beckwith}, S. V.~W., {et~al.} 2004, \apjs, 152,
  163

\bibitem[{{Roy} {et~al.}(2018){Roy}, {Napolitano}, {La Barbera}, {Tortora},
  {Getman}, {Radovich}, {Capaccioli}, {Brescia}, {Cavuoti}, {Longo}, {Raj},
  {Puddu}, {Covone}, {Amaro}, {Vellucci}, {Grado}, {Kuijken}, {Verdoes Kleijn},
  \& {Valentijn}}]{Roy2018}
{Roy}, N., {Napolitano}, N.~R., {La Barbera}, F., {et~al.} 2018, \mnras, 480,
  1057

\bibitem[{{Salim} {et~al.}(2007){Salim}, {Rich}, {Charlot}, {Brinchmann},
  {Johnson}, {Schiminovich}, {Seibert}, {Mallery}, {Heckman}, {Forster},
  {Friedman}, {Martin}, {Morrissey}, {Neff}, {Small}, {Wyder}, {Bianchi},
  {Donas}, {Lee}, {Madore}, {Milliard}, {Szalay}, {Welsh}, \&
  {Yi}}]{Salim2007ApJS..173..267S}
{Salim}, S., {Rich}, R.~M., {Charlot}, S., {et~al.} 2007, \apjs, 173, 267

\bibitem[{{S{\'a}nchez-Portal} {et~al.}(2015){S{\'a}nchez-Portal},
  {Pintos-Castro}, {P{\'e}rez-Mart{\'\i}nez}, {Cepa}, {P{\'e}rez Garc{\'\i}a},
  {Dom{\'\i}nguez-S{\'a}nchez}, {Bongiovanni}, {Serra}, {Alfaro}, {Altieri},
  {Arag{\'o}n-Salamanca}, {Balkowski}, {Biviano}, {Bremer}, {Castander},
  {Casta{\~n}eda}, {Castro-Rodr{\'\i}guez}, {Chies-Santos}, {Coia}, {Diaferio},
  {Duc}, {Ederoclite}, {Geach}, {Gonz{\'a}lez-Serrano}, {Haines}, {McBreen},
  {Metcalfe}, {Oteo}, {P{\'e}rez-Fourn{\'o}n}, {Poggianti}, {Polednikova},
  {Ram{\'o}n-P{\'e}rez}, {Rodr{\'\i}guez-Espinosa}, {Santos}, {Smail}, {Smith},
  {Temporin}, \& {Valtchanov}}]{glace2015A&A...578A..30S}
{S{\'a}nchez-Portal}, M., {Pintos-Castro}, I., {P{\'e}rez-Mart{\'\i}nez}, R.,
  {et~al.} 2015, \aap, 578, A30

\bibitem[{{Schawinski} {et~al.}(2014){Schawinski}, {Urry}, {Simmons},
  {Fortson}, {Kaviraj}, {Keel}, {Lintott}, {Masters}, {Nichol}, {Sarzi},
  {Skibba}, {Treister}, {Willett}, {Wong}, \&
  {Yi}}]{Schawinski2014MNRAS.440..889S}
{Schawinski}, K., {Urry}, C.~M., {Simmons}, B.~D., {et~al.} 2014, \mnras, 440,
  889

\bibitem[{{Scoville} {et~al.}(2007){Scoville}, {Aussel}, {Brusa}, {Capak},
  {Carollo}, {Elvis}, {Giavalisco}, {Guzzo}, {Hasinger}, {Impey}, {Kneib},
  {LeFevre}, {Lilly}, {Mobasher}, {Renzini}, {Rich}, {Sanders}, {Schinnerer},
  {Schminovich}, {Shopbell}, {Taniguchi}, \&
  {Tyson}}]{Cosmos2007ApJS..172....1S}
{Scoville}, N., {Aussel}, H., {Brusa}, M., {et~al.} 2007, \apjs, 172, 1

\bibitem[{{S\'ersic}(1968)}]{Sersic1968adga.book.....S}
{S\'ersic}, J.~L. 1968, {Atlas de Galaxias Australes} (Observatorio
  Astron\'omico, C\'ordoba, Argentia)

\bibitem[{{Shen} {et~al.}(2003){Shen}, {Mo}, {White}, {Blanton}, {Kauffmann},
  {Voges}, {Brinkmann}, \& {Csabai}}]{Shen2003MNRAS.343..978S}
{Shen}, S., {Mo}, H.~J., {White}, S. D.~M., {et~al.} 2003, \mnras, 343, 978

\bibitem[{{Simard}(1998)}]{Simard1998ASPC..145..108S}
{Simard}, L. 1998, Astronomical Society of the Pacific Conference Series, Vol.
  145, {GIM2D: an IRAF package for the Quantitative Morphology Analysis of
  Distant Galaxies} (ASPCS), 108

\bibitem[{{Simard} {et~al.}(2011){Simard}, {Mendel}, {Patton}, {Ellison}, \&
  {McConnachie}}]{Simard2011ApJS..196...11S}
{Simard}, L., {Mendel}, J.~T., {Patton}, D.~R., {Ellison}, S.~L., \&
  {McConnachie}, A.~W. 2011, \apjs, 196, 11

\bibitem[{{Simard} {et~al.}(2002){Simard}, {Willmer}, {Vogt}, {Sarajedini},
  {Phillips}, {Weiner}, {Koo}, {Im}, {Illingworth}, \&
  {Faber}}]{Simard2002ApJS..142....1S}
{Simard}, L., {Willmer}, C. N.~A., {Vogt}, N.~P., {et~al.} 2002, \apjs, 142, 1

\bibitem[{{Strateva} {et~al.}(2001){Strateva}, {Ivezi{\'c}}, {Knapp},
  {Narayanan}, {Strauss}, {Gunn}, {Lupton}, {Schlegel}, {Bahcall}, {Brinkmann},
  {Brunner}, {Budav{\'a}ri}, {Csabai}, {Castander}, {Doi}, {Fukugita},
  {Gy{\H{o}}ry}, {Hamabe}, {Hennessy}, {Ichikawa}, {Kunszt}, {Lamb}, {McKay},
  {Okamura}, {Racusin}, {Sekiguchi}, {Schneider}, {Shimasaku}, \&
  {York}}]{Strateva2001AJ....122.1861S}
{Strateva}, I., {Ivezi{\'c}}, {\v{Z}}., {Knapp}, G.~R., {et~al.} 2001, \aj,
  122, 1861

\bibitem[{{Tully} \& {Fisher}(1977)}]{TF_RELATION_1977A&A....54..661T}
{Tully}, R.~B. \& {Fisher}, J.~R. 1977, \aap, 500, 105

\bibitem[{{van der Wel} {et~al.}(2014){van der Wel}, {Franx}, {van Dokkum},
  {Skelton}, {Momcheva}, {Whitaker}, {Brammer}, {Bell}, {Rix}, {Wuyts},
  {Ferguson}, {Holden}, {Barro}, {Koekemoer}, {Chang}, {McGrath},
  {H{\"a}ussler}, {Dekel}, {Behroozi}, {Fumagalli}, {Leja}, {Lundgren},
  {Maseda}, {Nelson}, {Wake}, {Patel}, {Labb{\'e}}, {Faber}, {Grogin}, \&
  {Kocevski}}]{vanderWel2014ApJ...788...28V}
{van der Wel}, A., {Franx}, M., {van Dokkum}, P.~G., {et~al.} 2014, \apj, 788,
  28

\bibitem[{{Vika} {et~al.}(2015){Vika}, {Vulcani}, {Bamford}, {H{\"a}u{\ss}ler},
  \& {Rojas}}]{Vika_MegaMorph2015}
{Vika}, M., {Vulcani}, B., {Bamford}, S.~P., {H{\"a}u{\ss}ler}, B., \& {Rojas},
  A.~L. 2015, \aap, 577, A97

\bibitem[{{Vulcani} {et~al.}(2014){Vulcani}, {Bamford}, {H{\"a}u{\ss}ler},
  {Vika}, {Rojas}, {Agius}, {Baldry}, {Bauer}, {Brown}, {Driver}, {Graham},
  {Kelvin}, {Liske}, {Loveday}, {Popescu}, {Robotham}, \&
  {Tuffs}}]{Vulcani_MegaMorph_2014}
{Vulcani}, B., {Bamford}, S.~P., {H{\"a}u{\ss}ler}, B., {et~al.} 2014, \mnras,
  441, 1340

\bibitem[{{Wright}(2006)}]{Wright2006PASP..118.1711W}
{Wright}, E.~L. 2006, \pasp, 118, 1711

\bibitem[{{York} {et~al.}(2000){York}, {Adelman}, {Anderson}, {Anderson},
  {Annis}, {Bahcall}, {Bakken}, {Barkhouser}, {Bastian}, {Berman}, {Boroski},
  {Bracker}, {Briegel}, {Briggs}, {Brinkmann}, {Brunner}, {Burles}, {Carey},
  {Carr}, {Castander}, {Chen}, {Colestock}, {Connolly}, {Crocker}, {Csabai},
  {Czarapata}, {Davis}, {Doi}, {Dombeck}, {Eisenstein}, {Ellman}, {Elms},
  {Evans}, {Fan}, {Federwitz}, {Fiscelli}, {Friedman}, {Frieman}, {Fukugita},
  {Gillespie}, {Gunn}, {Gurbani}, {de Haas}, {Haldeman}, {Harris}, {Hayes},
  {Heckman}, {Hennessy}, {Hindsley}, {Holm}, {Holmgren}, {Huang}, {Hull},
  {Husby}, {Ichikawa}, {Ichikawa}, {Ivezi{\'c}}, {Kent}, {Kim}, {Kinney},
  {Klaene}, {Kleinman}, {Kleinman}, {Knapp}, {Korienek}, {Kron}, {Kunszt},
  {Lamb}, {Lee}, {Leger}, {Limmongkol}, {Lindenmeyer}, {Long}, {Loomis},
  {Loveday}, {Lucinio}, {Lupton}, {MacKinnon}, {Mannery}, {Mantsch}, {Margon},
  {McGehee}, {McKay}, {Meiksin}, {Merelli}, {Monet}, {Munn}, {Narayanan},
  {Nash}, {Neilsen}, {Neswold}, {Newberg}, {Nichol}, {Nicinski}, {Nonino},
  {Okada}, {Okamura}, {Ostriker}, {Owen}, {Pauls}, {Peoples}, {Peterson},
  {Petravick}, {Pier}, {Pope}, {Pordes}, {Prosapio}, {Rechenmacher}, {Quinn},
  {Richards}, {Richmond}, {Rivetta}, {Rockosi}, {Ruthmansdorfer}, {Sandford},
  {Schlegel}, {Schneider}, {Sekiguchi}, {Sergey}, {Shimasaku}, {Siegmund},
  {Smee}, {Smith}, {Snedden}, {Stone}, {Stoughton}, {Strauss}, {Stubbs},
  {SubbaRao}, {Szalay}, {Szapudi}, {Szokoly}, {Thakar}, {Tremonti}, {Tucker},
  {Uomoto}, {Vanden Berk}, {Vogeley}, {Waddell}, {Wang}, {Watanabe},
  {Weinberg}, {Yanny}, {Yasuda}, \& {SDSS
  Collaboration}}]{York2000AJ....120.1579Y}
{York}, D.~G., {Adelman}, J., {Anderson}, Jr., J.~E., {et~al.} 2000, \aj, 120,
  1579

\end{thebibliography}

\begin{appendix} 

\end{appendix}
\end{document}